\documentclass[12pt,letterpaper]{article}

\usepackage{graphicx,array}
\usepackage{url}
\usepackage{color}
\usepackage{latexsym}
\usepackage{amsthm}
\usepackage{amsmath}
\usepackage{amssymb}
\usepackage{amsfonts}
\usepackage[numbers,sort&compress]{natbib}
\usepackage{bm}
\usepackage{slashed}
\usepackage{mathrsfs}
\usepackage{enumerate}
\usepackage{tikz}
\usepackage{siunitx}
\usepackage{mdframed}
\usepackage{setspace}  
\usepackage{esvect}
\usepackage{esint}
\usepackage{subcaption}
\usepackage[normalem]{ulem}
\usepackage{setspace}
\usepackage{multirow,makecell}

\usepackage[utf8x]{inputenc}%
\usepackage{tcolorbox}%

\usepackage{hyperref} 

\hypersetup{
    colorlinks=true,       
    linkcolor=red,          
    citecolor=blue,        
    filecolor=magenta,      
    urlcolor=blue           
}
\usepackage[all]{hypcap} 

\usepackage{natbib}
\newcommand{\nc}{\newcommand}

\nc{\beq}{\begin{equation}}  
\nc{\eeq}{\end{equation}}  
\nc{\beqa}{\begin{eqnarray}}  
\nc{\eeqa}{\end{eqnarray}}  
\nc{\bit}{\begin{itemize}}  
\nc{\eit}{\end{itemize}}

\usepackage{floatrow}
\newfloatcommand{capbtabbox}{table}[][\FBwidth]

\usepackage{blindtext}

\title{ 
{\bf More Scalings from Cosmic Strings
}
\author{\large Heejoo Kim$^{\,\dagger}$ and Minho Son$\,^\dagger$}
\date{\small \it 
$^\dagger$Department of Physics, Korea Advanced Institute of Science and Technology, \\
291 Daehak-ro, Yuseong-gu, Daejeon 34141, Republic of Korea\\
}
}

\begin{document}

\maketitle

\setlength{\parskip}{0.2ex}

\begin{abstract}	
We analyze all individual cosmic strings of various lengths in a large ensemble of the global cosmic string networks in the post-inflationary scenario, obtained from numerical simulations on a discrete lattice with $N^3 = 4096^3$. A strong evidence for a logarithmically growing spectral index of the string power spectrum during the evolution is newly reported as our main result. The logarithmic scaling is checked against two different approaches for generating initial random field configurations, namely fat-string type and thermal phase transition. We derive the analytic relation between two power spectra of cosmic strings and axions which should be valid under some assumptions, and the validity of those assumptions is discussed. We argue that our analytic result strongly supports the correlated spectra of cosmic strings and axions. Additionally, we initiate the statistical analysis of the causal dynamics of the cosmic strings.
\end{abstract}

\thispagestyle{empty}  
\newpage  
  
\setcounter{page}{1}

\begingroup
\hypersetup{linkcolor=black,linktocpage}
\tableofcontents
\endgroup
\newpage


\section{Introduction}
\label{sec:intro}

The QCD axion is a strong candidate not only for solving the mysterious strong CP problem in the Standard Model (SM), but also for accounting for the observed dark matter abundance~\cite{Peccei:1977hh,Weinberg:1977ma,Wilczek:1977pj,Dine:1982ah,Preskill:1982cy,Abbott:1982af}.
A typical mechanism for generating axions is through the spontaneous breaking of a global Pecci-Quinn (PQ) symmetry~\cite{Peccei:1977hh}. When the PQ symmetry is broken after the inflation, namely the post-inflationary scenario, the QCD axion abundance can be predicted in terms of only the axion mass. If the misalignment mechanism accounts for the entire observed value, the axion mass is predicted to be 
$m_a = 5.7 \mu{\rm eV} (10^{12}\, {\rm GeV}/f_a)$~\cite{GrillidiCortona:2015jxo} where $f_a$ is the axion decay constant. However, topological cosmic strings are inevitably generated upon the PQ phase transition~\cite{Kibble:1984hp}  and they can contribute to the dark matter by radiating into axions. A large contribution from cosmic strings relative to the misalignment increases the favored axion mass (thus decreases the axion decay constant). Any effort for pinning down the axion mass within the minimal QCD axion cosmology as precise as possible should be not only highly beneficial for current and future axion particle search programs~\cite{AxionLimits,ADMX:2003rdr,Caldwell:2016dcw,Petrakou:2017epq,CAST:2017uph,IAXO:2019mpb,ADMX:2019uok,Wuensch:1989sa,Hagmann:1990tj,Raffelt:2006cw,Viaux:2013lha,Arvanitaki:2014dfa,Ayala:2014pea,MillerBertolami:2014rka,Graham:2015ouw,Kahn:2016aff,Barbieri:2016vwg,Brubaker:2016ktl,Irastorza:2018dyq,Straniero:2018fbv,Chang:2018rso,Melcon:2018dba,Marsh:2018dlj,Lawson:2019brd,Carenza:2019pxu,Zarei:2019sva,Berlin:2019ahk,Lasenby:2019prg,Beurthey:2020yuq,Mitridate:2020kly,Capozzi:2020cbu,Kim:2021eye,Buschmann:2021juv,Foster:2022fxn,Escudero:2023vgv,Caputo:2024oqc,Khelashvili:2024sup,Manzari:2024jns,Hardy:2024fen,Garcia:2024xzc} but also establishing the late time QCD axion cosmology~\cite{Marsh:2015xka,OHare:2024nmr}.

The evolution of cosmic strings can reliably proceed only through numerical lattice simulations due to nonlinearities of interactions.
While the evolution should cover the entire dynamic time range until the time for the QCD crossover,
 $\log(m_r/H) \sim 70$ ($m_r^{-1}$ as the string core size, $H$ as the Hubble parameter), around which the axion potential becomes relevant and the string-domain wall network (that subsequently collapse into axions~\cite{Hiramatsu:2012sc,Hiramatsu:2012gg,Buschmann:2019icd,OHare:2021zrq}) forms~\cite{PhysRevD.45.3394,Sikivie:2006ni}, the dynamic time range on simulations is limited to roughly $\log(m_r/H) \sim 8-10$ and the rest of the time range relies on a huge extrapolation. Although it has been known that the cosmic string network eventually enters the scaling regime at late times, a limited dynamic time range makes an estimation of the contribution from topological strings sensitive to initial conditions and simulation setups.

Recently, axion dark matter from topological strings has been actively studied by several groups using the state-of-the-art numerical simulations on much larger grids than early studies in~\cite{Yamaguchi:1998gx,Yamaguchi:1999yp,Yamaguchi:1999dy,Yamaguchi:2002sh,Hiramatsu:2010yu,Kawasaki:2014sqa} (see related discussions in~\cite{Bennett:1985qt,Bennett:1986zn,Battye:1993jv,Battye:1994au,Martins:1995tg}) or using an advanced adaptive mesh refinement (AMR) technique~\cite{BERGER1984484} (AMReX~\cite{zhang2020amrex,AMReX_JOSS} and GRChombo~\cite{Clough:2015sqa} are available softwares and they were used for the study of cosmic strings in~\cite{Drew:2019mzc,Buschmann:2021sdq,Benabou:2023ghl,Buschmann:2024bfj}). The increased dynamic time range covered by recent lattice simulations have revealed many intriguing observations, not only strong evidences for logarithmic scalings of the number of strings per Hubble patch~\cite{Gorghetto:2018myk,Kawasaki:2018bzv,Vaquero:2018tib,Klaer:2019fxc,Buschmann:2019icd,Gorghetto:2020qws,Buschmann:2021sdq,OHare:2021zrq,Saikawa:2024bta,Kim:2024wku} and the spectral index of the axion spectra~\cite{Gorghetto:2020qws,Saikawa:2024bta,Kim:2024wku}, but also the relevance of a few previously overlooked epochs of nonlinearities after the time for the QCD phase transition~\cite{Gorghetto:2020qws,Gorghetto:2024vnp}, modifying our understanding of the QCD axion cosmology (see~\cite{Vaquero:2018tib,Eggemeier:2019khm,OHare:2021zrq,Xiao:2021nkb,Shen:2022ltx,Eggemeier:2022hqa,Gorghetto:2022ikz,Pierobon:2023ozb,Gorghetto:2023vqu} for early studies). 
The axion spectrum should follow the power law fall-off scaling like $k^{-q}$, where the spectral index $q$ characterizes the nature of axions, in between widely separated two scales of the PQ symmetry breaking and $H$ around the QCD crossover.
A logarithmically growing spectral index $q$ indicates that QCD axions radiated from strings around the time for the QCD phase transition are IR-dominated and, as a result, an enhanced abundance for axions radiated from topological strings can fully account for the current dark matter abundance.
Taking into account for the effect due to nonlinearities from the axion potential, the axion mass is predicted to be as large as 400$\sim$500 $ \mu$eV and the axion decay constant as low as $f_a \sim 10^{10}$ GeV~\cite{Gorghetto:2020qws,Kim:2024wku}~\footnote{Different fit result, e.g. the spectral index $q$ of the order of the unity, of a similar ansatz and different treatment of nonlinearities around QCD crossover can change the prediction, $m_a = [40,\, 180]\mu eV$~\cite{Buschmann:2021sdq}. Various combinations of ansatz for fitting the number of strings per Hubble patch and the axion spectrum predict a broad range of the axion mass, 
$m_a = [95,\, 450] \mu eV$~\cite{Saikawa:2024bta}. The axion mass window in~\cite{Buschmann:2021sdq,Saikawa:2024bta} amounts to the uncertainty of the numerical result and its analysis while the axion dark matter abundance matches to the full observed value. Whereas the axion masses in~\cite{Gorghetto:2020qws,Kim:2024wku} are lower bounds for a choice of parameters by demanding the axion abundance does not exceed the observed value.}. However, numerical results supporting for an order one constant scaling of the string population per Hubble patch or the axion spectrum with the spectral index of a unity also exist~\cite{Hindmarsh:2019csc,Buschmann:2021sdq,Hindmarsh:2021zkt,Correia:2024cpk}. A follow-up study will be necessary to resolve the discrepancy.

In this work, we take a different avenue to understand the logarithmic scaling observed in the axion spectrum. Since axions are radiated dominantly from the cosmic strings with a sub-dominant contribution from radial modes~\cite{Gorghetto:2018myk,Benabou:2023ghl,Kim:2024wku}, the characteristic feature of the axion spectrum should be closely related to the spectrum of strings themselves. Along this line of reasoning, we analyze fluctuations of the full set of strings of various lengths in the cosmic string network. Similarly to the case of the axion spectrum, we newly observe a logarithmically growing spectral index of the string power spectrum. This is the main result of this work. 
In a situation where an almost straight string is a valid approximation, provided the topological interaction between axions and strings in the thin string limit~\cite{Kalb:1974yc}, we demonstrate that the axion power spectrum can be analytically derived from the string fluctuations. Although the truth relation between two spectra of axions and strings looks subject to the contribution from the region beyond the assumptions that we made in our analytic derivation, we argue that our derivation strongly supports the correlated logarithmic growths observed in both axion and string spectra.

Our paper is organized as follows. 
In Section~\ref{sec:cosmic:strings}, the formation of cosmic strings and its simulation setup on the lattice are briefly discussed. A detailed discussion on string loop distributions from our lattice simulations is given.
In Section~\ref{sec:stringPS}, the description of our approach for studying the string power spectrum is given. We present our result of the string power spectrum exhibiting the power-law scaling behavior from which the spectral index of string spectrum is extracted. 
In Section~\ref{sec:strings:axions}, our derivation of the analytic relation between two spectral indices of cosmic strings and axions is given and the validity of it is discussed. 
In Appendix~\ref{app:setups}, a further description for our simulation setup is provided. Especially, two relaxation schemes to boost the approach of the string network toward the attractor solution are explained.
In Appendix~\ref{app:sec:morestring}, more supplementary plots for the string spectrum are presented.
In Appendix~\ref{sec:cs:randomwalks}, our exercise on the cosmic random walk strings is presented and compared with the string spectrum in the region of $k<H$. 
In Section~\ref{sec:dyn:string}, we initiate the statistical analysis of the causal dynamics of cosmic strings. For the purpose of the illustration, we present the detailed statistical distributions of the single string evolutions (the causal string dynamics that do not go through a merging or a branching). We show that the corresponding statistical distributions provide a quantitative measure for the validity of our aforementioned analytic result.

\section{Cosmic string network}
\label{sec:cosmic:strings}

\subsection{Formation and simulation setup}

The spontaneous breaking of the PQ symmetry is realized by the following potential of the complex scalar $\phi$,
\begin{equation}\label{eq:Lag:orig}
  \mathcal{L} = \partial_\mu \phi ^* \partial^\mu \phi - \frac{m_r^2}{2 f_a^2} \left ( |\phi|^2 - \frac{f_a^2}{2} \right )^2~,
\end{equation}
where $f_a$ is the PQ symmetry breaking scale and $m_r$ is the mass of the radial fluctuation. 
The equation of motion from the Lagrangian in Eq.~\ref{eq:Lag:orig}, during the radiation dominated era in the expanding universe with the scale factor of $R(t) \propto \sqrt{t}$, is given by
\begin{equation}\label{eq:phi:eom}
  \ddot{\phi} + 3 H \dot{\phi} - \frac{1}{R^2} \nabla ^2 \phi + \frac{m_r^2}{f_a^2} \phi \left ( |\phi|^2 - \frac{f_a^2}{2} \right ) = 0~,
\end{equation}
where the dot and the gradient $\nabla$ denote differentiations with respect to the cosmic time $t$ and comoving coordinates, respectively. 
The equation of motion admits a solitonic solution for a topological cosmic string~\cite{Kibble:1976sj,Kibble:1980mv,Vilenkin:1981kz}. The dynamic time range that can be simulated on the lattice is severely limited due to different scalings of $m_r^{-1}$, $H^{-1}$, and $R(t)$ over time, and the maximum dynamic time is given by roughly $\log(m_r/H) \leq \log [N/(n_c n_H)]$ where $N$ is the lattice size per dimension and $n_c$ ($n_H$) is the number of lattice points within the string core (the number of Hubble patches in the simulation box) per dimension at the final time. The values of $n_c=1$ and $n_H = 4^{1/3}$ are chosen in our simulation.

The discretization of the equation of motion in Eq.~\ref{eq:phi:eom} and its evolution on the lattice were carried out with the same simulation setup as in~\cite{Kim:2024wku}. Our data set contains newly generated 100 independent simulations on a lattice with $N^3=4096^3$ where initial conditions for the physical string evolution were prepared by the fat-string pre-evolution and thermal pre-evolution (see Appendix~\ref{app:setups} for the description). 
As was in~\cite{Kim:2024wku}, $\xi_0 = 0.2$ for the number of strings per Hubble patch at the initial time of the physical string evolution, or $\xi_0 (\log (m_r/H)=2) = 0.2$, was chosen as the benchmark scenario which is believed to be close to the truth attractor solution. Similarly, as was in~\cite{Kim:2024wku}, $\zeta = 2.12$ was chosen as the benchmark scenario when the thermal pre-evolution was used. 
In the simulation approach using the fat-string pre-evolution during which $\frac{m_r^{-1}}{R}$ remains constant, the field configuration after the fat-string evolution is taken as the initial condition for the physical string evolution. 
The choice of the scale factor $R\propto t$ during the fat-string evolution~\cite{Gorghetto:2020qws,Saikawa:2024bta,Kim:2024wku} ensures that the correlation length in the string core width unit stays constant, and it can be freely chosen.
In the approach using the thermal pre-evolution, the Lagrangian in Eq.~\ref{eq:Lag:orig} is supplemented by the thermal potential $\Delta \mathcal{L}= - \frac{m_r^2}{6f_a^2} T^2 |\phi|^2$~\cite{Kawasaki:2018bzv,Buschmann:2021sdq,Kim:2024wku} where the temperature $T$ is related to $\zeta$, that is used to select an appropriate initial condition, and the Hubble parameter $H$ through the relation $T^4 = f_a^4\cdot \frac{H^2}{m_r^2}\cdot 4 \zeta^2$ where $H^2 = \frac{\pi^2}{90} g_* \frac{T^4}{M_p^2}$, $\zeta^2 = \frac{45}{2\pi^2 g_*}\frac{M_p^2 m_r^2}{f_a^4}$, and $M_p$ is the Planck mass.
The equation of motion is evolved from an initial time before the phase transition for the spontaneous breaking of the PQ symmetry.  The correlation length in the string core width unit at the critical temperature $T_c$ is given by $\frac{H^{-1}}{m_r^{-1}} |_{T_c}=\frac{2}{3}\zeta$.
In both approaches, string identifications were performed at time slices with the time step of $\pi m_r^{-1}$, corresponding to half the period of the radial mode oscillation below the $m_r$ scale, for the study of the causal dynamics of the cosmic string network.

\subsection{String distribution}

A string is identified when a nontrivial pattern of the phases of the complex fields is found. In our lattice simulations, strings were identified with the tetrahedron-based algorithm, first introduced in~\cite{Kim:2024wku}, which guarantees the connectedness of strings and provides a systematic way of the string core identification. Other types of string identifications that ensure the connectedness of strings can be found in~\cite{Yamaguchi:2002zv,Yamaguchi:2002sh,Hiramatsu:2010yu}.

\begin{figure}[tp]
\begin{center}
\includegraphics[width=0.55\textwidth]{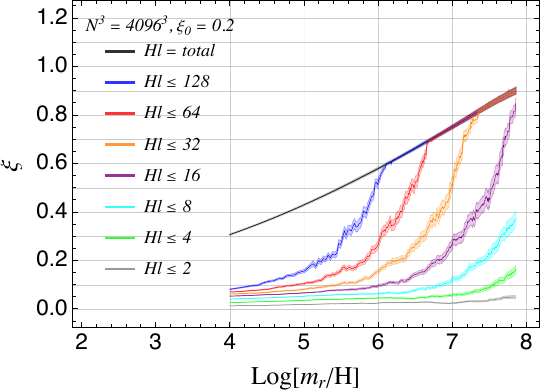}
\caption{\small The number of strings per Hubble patch $\xi$ from the simulation with the initial condition $\xi_0 (\log\frac{m_r}{H} = 2) = 0.2$. The curves after $\log\frac{m_r}{H} = 4$ are shown. The total contribution to $\xi$ is decomposed into individual ones of various string lengths.
}
\label{fig:xi:composition:fat:4096}
\end{center}
\end{figure}

Cosmic strings decay through intercommutations~\cite{Moore:2016itg} and evaporations during the expansion of the universe.
A characteristic property of such dynamics can be captured by measuring the number of strings per Hubble patch, denoted by $\xi$. While the decay of strings and the expansion of the universe compete each other in the contribution to $\xi$, the evolution of $\xi$ exhibits a strong evidence for the logarithmically growing scaling solution at late times (see Fig.~\ref{fig:xi:composition:fat:4096}), irrespective of initial conditions for the cosmic string network~\cite{Gorghetto:2018myk,Gorghetto:2020qws,Buschmann:2021sdq,Saikawa:2024bta,Kim:2024wku}.

\begin{figure}[tp]
\begin{center}
\includegraphics[width=0.55\textwidth]{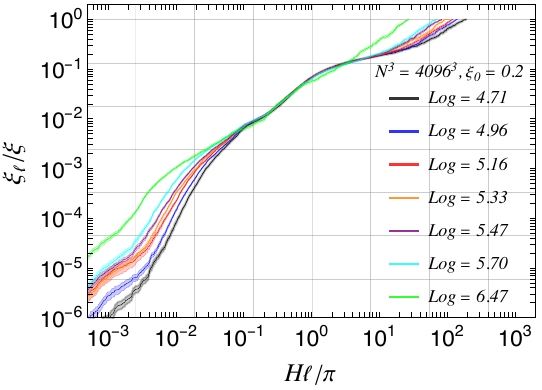}
\caption{\small The fraction of string loops whose lengths are smaller than $\ell$, contributing to $\xi$, for selected time slices. The curves are based on the simulation with the initial condition $\xi_0 (\log\frac{m_r}{H} = 2) = 0.2$. 
}
\label{fig:loopstrings:fat:4096}
\end{center}
\end{figure}

We sort all the string loops in the string network according to their lengths and they are grouped into several bins of the string length for the analysis of the string power spectrum later.
Before doing it, similarly to~\cite{Gorghetto:2018myk,Buschmann:2021sdq}~\footnote{The discussion in~\cite{Gorghetto:2018myk} was based on the analysis of fat-strings, as opposed to our physical strings, although some properties are shared.}, 
we can look into the composition of string loops in $\xi$.
The string loop density whose lengths are shorter than $\ell$ can be defined as
\begin{equation}
\xi_\ell  = \frac{t^2}{L^3}\sum_{s=strings} \ell_s \theta(\ell - \ell_s)~,
\end{equation}
where sum is done over strings of length $\ell_s \leq \ell$ contained in the simulation box of the size $L$. In the limit of $\ell \rightarrow \infty$, the summation becomes total string length in the simulation box, $\ell_\text{tot}(L)$, and the string loop density asymptotes to $\xi_\infty = \xi = \ell_\text{tot}(L) t^2/L^3$, the number of strings per Hubble patch. Therefore, $\xi_\ell/\xi$ represents the fractional contribution of string loops shorter than $\ell$ to $\xi$, and our simulation result is illustrated in Fig.~\ref{fig:loopstrings:fat:4096}.

A lesson from Fig.~\ref{fig:loopstrings:fat:4096} is that the population of sub-Hubble strings of $\ell \lesssim H^{-1}$ is less than 10\% and the dominant contribution to $\xi$ comes from super-Hubble strings of $\ell \gtrsim \mathcal{O}(10^2)\, H^{-1}$ for time slices shown in Fig.~\ref{fig:loopstrings:fat:4096}. Those super-Hubble strings contributing to $\xi$ the most are long strings that wrap the entire simulation box multiple times. 
Recall that a long string of a length larger than the simulation box size, $\ell \gg L = \pi H^{-1} (LH/\pi)$, where the number of Hubble patches per dimension is given by $L/H^{-1}= \sqrt{ n_H N/(n_c e^{\log(m_r/H)} ) }$ and $L/H^{-1} \sim 4$ at $\log(m_r/H) = 6$ for our choice of $n_H$ and $n_c$, necessarily wraps the simulation box multiple times.
The role of long strings can be more directly seen in Fig~\ref{fig:xi:composition:fat:4096} where the contribution to $\xi$ was sub-divided in terms of string lengths.
As is evident in Figs~\ref{fig:loopstrings:fat:4096} and~\ref{fig:xi:composition:fat:4096}, in an early time with a small Hubble length, $\xi$ obtains the largest contribution from long strings of the largest available $H\ell$, for example, even long strings of $H\ell \lesssim 128$ take only tiny portion of $\xi$ around $\log(m_r/H) \sim 4$. The length of long strings in Hubble unit, accounting for almost the entire portion of the string population in $\xi$,  becomes smaller and smaller over the time due to the increasing Hubble length. 

In Fig.~\ref{fig:loopstrings:fat:4096}, the fraction of string loops of order of the length $\ell \sim \pi H^{-1}[10^{-1},\,10]$ stays constant over time which implies that both $\xi_\ell$ and $\xi$ have similar logarithmic scalings in the corresponding range of string lengths. Indeed, their similar scaling behaviors can be directly seen in Fig.~\ref{fig:xi:composition:fat:4096}. It indicates that the rate of the immigration of strings to $\xi_\ell$ (for a given $\ell$) is balanced with the migration rate of strings out of $\xi_\ell$ for strings in this range of lengths. 
Once strings belonging to $\xi_\ell$ exit from this scaling regime at late times, their contributions to $\xi$ rapidly increase to get saturated.

On the other hand, the string loops of lengths shorter than the wavelength of the string core size, $\ell \lesssim 2\pi m_r^{-1}$, should belong to the regime that is sensitive to the UV modeling. The range can be re-written as $\ell \lesssim 2\pi H^{-1}/e^{\log(m_r/H)}$ from which $\ell \lesssim \pi H^{-1} [0.3,\, 2] \times 10^{-2}$ for $\log(m_r/H) = [4.7,\, 6.7]$ (the largest and smallest time slices in Fig.~\ref{fig:loopstrings:fat:4096}). While $\xi_\ell/\xi$ in this regime grows over time, as is seen in Fig.~\ref{fig:loopstrings:fat:4096}, the relative contribution to $\xi$ is negligible.

Based on observations discussed in this section, the axions radiated from super-Hubble strings of $\ell \gtrsim H^{-1}$ should be relevant ones that will determine the abundance of the axion dark matter. 
Since the dominant contribution comes from long strings in the string network, the characteristic feature of the power spectrum of those long strings may well match to that of the axion power spectrum. This will be discussed below in detail.

\section{String power spectrum}
\label{sec:stringPS}

\subsection{Parametrization of string fluctuation}

\begin{figure}[tp]
\begin{center}
\includegraphics[width=0.48\textwidth]{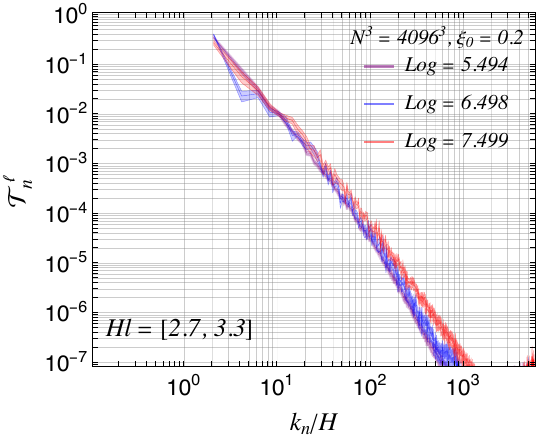}\quad
\includegraphics[width=0.48\textwidth]{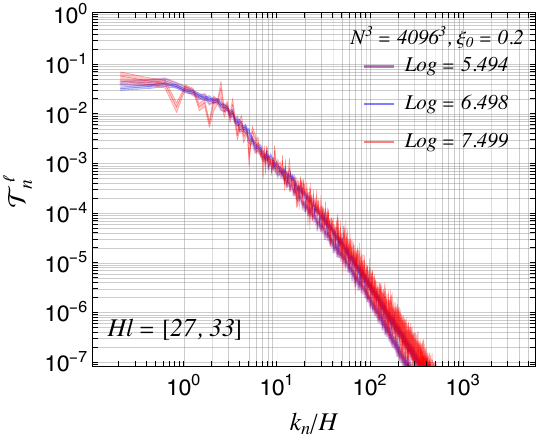}
\\[5pt]
\includegraphics[width=0.48\textwidth]{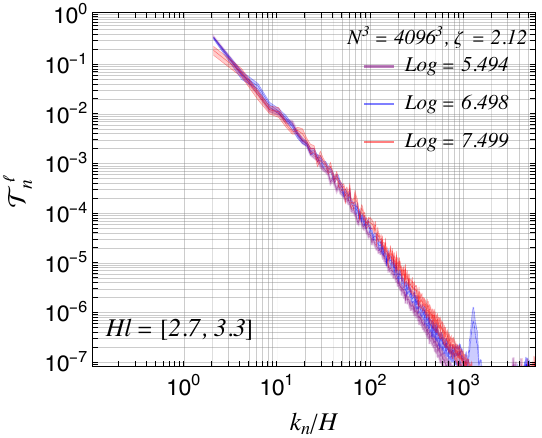}\quad
\includegraphics[width=0.48\textwidth]{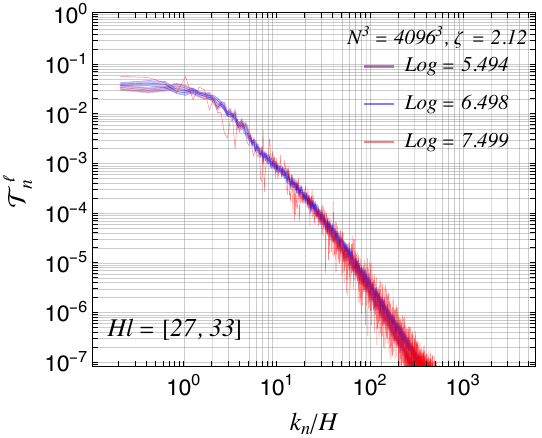}
\caption{\small The string power spectrum $\mathcal{T}_n^\ell$ for cosmic strings of lengths $\ell = [2.7,\, 3.3] H^{-1}$ and $[27,\, 33] H^{-1}$ at three time slices. Top panels were based on data assisted by the fat-string pre-evolution for the benchmark scenario of $\xi_0 = 0.2$. Bottom panels were based on data assisted by the thermal pre-evolution for the benchmark scenario of $\zeta = 2.12$.
}
\label{fig:strPS:4096}
\end{center}
\end{figure}

A string configuration can be parametrized in terms of a single variable $s$, namely $\vec{\gamma}(s)$, and a tangent vector $\vec{t}$ at $s$ along the string of a length $\ell$ can be written as
\begin{equation}
\frac{d\vec{\gamma}(s)}{ds} = \vec{t} (s) 
= \sum_{n=-\infty}^{\infty} \vec{T}^{\ell}_n e^{i \frac{2\pi n}{\ell} s}~,
\end{equation}
where the Fourier amplitude $\vec{T}^\ell_n$ quantifies magnitudes of fluctuations at different length scales on a string and it is obtained by the inverse transform,
\begin{equation}\label{eq:T:Fourieramp}
  \vec{T}^\ell_n = \int_0^\ell \frac{ds}{\ell} \vec{t}(s) e^{-i\frac{2\pi n}{\ell} s}~.
\end{equation}
A random string configuration of a length $\ell$ corresponds to the random sampling of amplitudes $\vec{T}^\ell_n$. 
We measure the statistical properties of those amplitudes out of the cosmic string network obtained from our lattice simulations through the 2-point function that can be defined as
\begin{equation}
  \langle \vec{T}^\ell_{n}  \cdot \vec{T}^\ell_{n'} \rangle = \mathcal{T}^\ell_{|n|} \delta_{(n+n')0}~,
\end{equation}
where the form of the 2-point function was constrained by the invariance under a shift of $s$ by a constant and an inversion of $s$. Since a tangent vector $\vec{t}$ is real, the amplitude satisfies $(\vec{T}^\ell_n)^* = \vec{T}^\ell_{-n}$ which implies that 
$\mathcal{T}^\ell_{|n|} = \langle |\vec{T}^\ell_{n}|^2 \rangle$ is real. Therefore, the string spectrum of the 2-point function can be characterized by a real function $\mathcal{T}^\ell_{n}$ for non-negative values of $n$, or $n \geq 0$. The amplitudes of different modes are constrained to satisfy the normalization,
\begin{equation}\label{eq:Tn:norm}
  1 = |\vec{t}(s)|^2 = \sum_{n=-\infty}^{\infty} \sum_{n'=-\infty}^{\infty} \vec{T}^\ell_{n} \cdot \vec{T}^\ell_{n'} e^{i\frac{2\pi (n+n')}{\ell} s}~,
\end{equation}
for each value of $s$. Upon taking an ensemble average, Eq.~\ref{eq:Tn:norm} implies the normalization of the string spectrum function $\mathcal{T}^\ell_n$, namely
\begin{equation}
  1 = \sum_{n=-\infty}^{\infty} \mathcal{T}^\ell_n~.
\end{equation}

Since a cosmic string network from each lattice simulation consists of strings of different lengths, strings are binned according to their lengths and an ensemble average is taken over them within each bin. Strings of sub-Hubble and super-Hubble lengths will be treated separately. For strings of sub-Hubble length, $0<\ell < H^{-1}$, we divide the region into bins of a size $\Delta\ell \approx 2\pi m_r^{-1}$, or a wavelength of the string core scale.  Strings of super-Hubble length, or $\ell  > H^{-1}$, are grouped into small bins of sizes $(1.1)^n H^{-1} < \ell \leq (1.1)^{n+1} H^{-1}$ with non-negative integer $n$. Multiple bins are merged into a single bin later to include strings in an appropriate range of a string length.
Our results for the string spectrum function $\mathcal{T}^\ell_n$ for two benchmark string lengths, $\ell \sim 3H^{-1}$ and $30H^{-1}$ with 10\% of widths, are shown in Fig.~\ref{fig:strPS:4096} which indeed show the power law fall-off scaling. String fluctuations of larger lengths than $H^{-1}$, or the modes in the region of $k_n/H \lesssim 1$, in the right panel of Fig.~\ref{fig:strPS:4096} will be frozen in the expanding universe, and thus the corresponding modes do not contribute to the axion spectrum. On the other hand, those fluctuations of super-Hubble wavelengths in the region of $k_n/H \lesssim 1$ are those that can be well-captured by a random walk with a constant step size of $H^{-1}$ (see Appendix~\ref{sec:cs:randomwalks} along with Fig.~\ref{fig:randomWalk:4096} for this exercise that shows a good agreement).

\subsection{Power-law scaling of string fluctuation}

Assuming that axion spectra are closely related to those of strings, the string spectra are expected to follow the power-law fall off behavior between the PQ symmetry breaking and Hubble scales, that is $\mathcal{T} \propto k^{-p}$. Similarly to the situation for the axion spectrum, contaminations from UV and IR regions can be reduced by imposing appropriate UV/IR cuts on the momentum, namely
$x^\text{str}_\text{IR} < k/H < y/x^\text{str}_\text{UV}$ where $y=m_r/H$ and the superscript `str' was to distinguish the cuts from those $x_\text{IR}$ and $x_\text{UV}$ used in the axion spectrum~\cite{Gorghetto:2020qws,Kim:2024wku}. The proper choice of cuts will be chosen based on our empirical observation against what can be suggested from an analytic estimation in a limited situation~\footnote{While a rough guideline for cuts may be given by noting that the energy from a sinusoidal string vibration of a wavelength $\lambda$ is mostly radiated into axions of wavelength $\lambda/2$~\cite{Battye:1993jv,Drew:2019mzc}, this property is valid only for a sinusoidal oscillation of a single frequency. 
An educated choice of the cuts is derived in Section~\ref{app:sec:strings:to:axions} although the suggestion should work under certain assumptions.}.

While only a collective contribution to the axion spectrum is examined in the region away from strings, contributions to the string spectrum from strings of different lengths can be separately examined. Since a cosmic string can be thought of as a three-dimensional self-avoiding random walks whose correlation length is of an order of $H^{-1}$, and we are interested in the string spectrum in a length scale shorter than the Hubble length, we suspect that the power law fall-off behavior should be similar for strings of $\ell \gg H^{-1}$. For instance, similar spectral indices are expected as the string power spectrum is averaged over a large $\ell/H^{-1}$ number of Hubble patches and thus it should represent a well-behaved statistical property. Whereas strings of sub-Hubble lengths may show different power-law behaviors and suffer from small statistics. Since the spacing of the rescaled momentum is given by $\Delta (k/H)|_\text{str} = 2\pi/(H\ell)$ for a string of a length $\ell$, the number of sampling points used for the fit within the interval is $(y/x^\text{str}_\text{UV} - x^\text{str}_\text{IR})H\ell/(2\pi)$. Recall that the spacing of the rescaled momentum in the axion power spectrum is $\Delta (k/H)|_\text{axion} = 2^{3/2}\pi \sqrt{y} (n_c/(2n_H N))^{1/2}$~\cite{Kim:2024wku} which corresponds to a string length of an order of the simulation box, $\ell \sim R N\Delta x$ ($\Delta x$ as the lattice spacing). For strings whose lengths are shorter than the simulation box size at a given time slice, $\Delta (k/H)|_\text{str} > \Delta (k/H)|_\text{axion}$, or smaller number of sampling points than the case for the axion spectrum, is expected, assuming the same sized fit interval. In our fitting procedure, we demand at least 10 sampling points for a reliable fit.

\begin{figure}[tp]
\begin{center}
\includegraphics[width=0.48\textwidth]{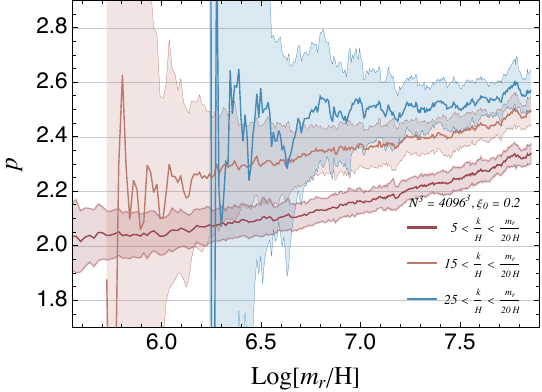}\quad
\includegraphics[width=0.48\textwidth]{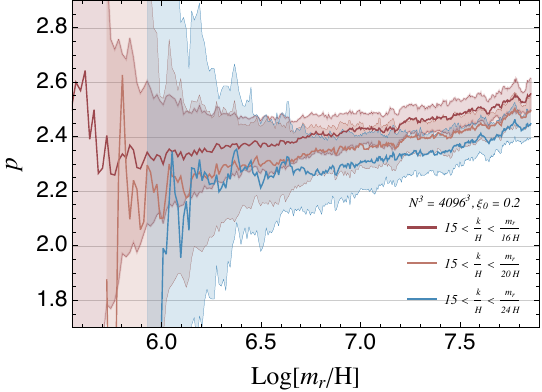}
\caption{\small The averaged spectral index $p$ of the string power spectrum for super-Hubble strings of $\ell > 5H^{-1}$ obtained by fitting the data within the momentum interval, $x^\text{str}_\text{IR} < k/H < m_r/(x^\text{str}_\text{UV}H)$. In the left (right) panel, $x^\text{str}_\text{IR}$ ($x^\text{str}_\text{UV}$) is varied while $x^\text{str}_\text{UV}$ ($x^\text{str}_\text{IR}$) is fixed. Both panels were based on data assisted by the fat-string pre-evolution for the benchmark scenario of $\xi_0 = 0.2$.
}
\label{fig:p:Lgtr5H:xUV:xIR:varied}
\end{center}
\end{figure}

\begin{figure}[tp]
\begin{center}
\includegraphics[width=0.48\textwidth]{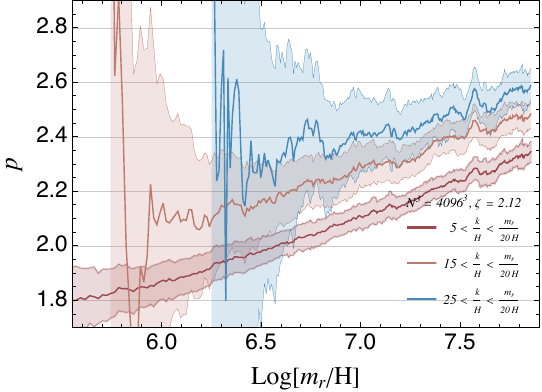}\quad
\includegraphics[width=0.48\textwidth]{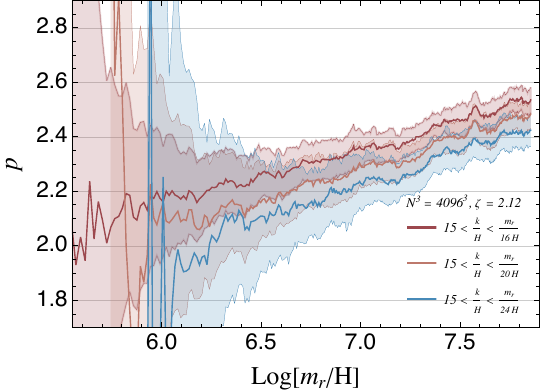}
\caption{\small The averaged spectral index $p$ of the string power spectrum for super-Hubble strings of $\ell > 5H^{-1}$ obtained by fitting the data within the momentum interval, $x^\text{str}_\text{IR} < k/H < m_r/(x^\text{str}_\text{UV}H)$. In the left (right) panel, $x^\text{str}_\text{IR}$ ($x^\text{str}_\text{UV}$) is varied while $x^\text{str}_\text{UV}$ ($x^\text{str}_\text{IR}$) is fixed. Both panels were based on data assisted by the thermal pre-evolution for the benchmark scenario of $\zeta = 2.12$.
}
\label{fig:p:Lgtr5H:xUV:xIR:varied:thermal}
\end{center}
\end{figure}

Assuming that strings of $\ell \gtrsim 5 H^{-1}$ share similar spectral indices (see Fig.~\ref{fig:strp:4096:fixedHL} for our numerical confirmation), we estimate the averaged spectral index as
\begin{equation}\label{eq:ave:p}
 p = \frac{1}{\Delta \ell} \int_{\ell_\text{min}}^{\ell_\text{max}} d\ell p(\ell)~,
\end{equation}
where $\Delta\ell = \ell_\text{max} - \ell_\text{min}$ with $\ell_\text{min}= 5H^{-1}$ and $p(\ell)$ is the spectral index fitted for a string of a length $\ell$.
The evolutions of the spectral indices, defined as Eq.~\ref{eq:ave:p}, are illustrated in Figs.~\ref{fig:p:Lgtr5H:xUV:xIR:varied} and~\ref{fig:p:Lgtr5H:xUV:xIR:varied:thermal} for two benchmark scenarios with $\xi_0 = 0.2$ and $\zeta = 2.12$, respectively. In the left panels of Figs.~\ref{fig:p:Lgtr5H:xUV:xIR:varied} and~\ref{fig:p:Lgtr5H:xUV:xIR:varied:thermal}, the spectral indices of string fluctuations are extracted using the interval with the varying $x^\text{str}_\text{IR}$ cut whereas, in the right panels, only $x^\text{str}_\text{UV}$ cut was varied. 
A clear logarithmic growth of the spectral index $p$ for strings of super-Hubble length, given appropriate UV and IR cuts, is observed in both cases of Figs.~\ref{fig:p:Lgtr5H:xUV:xIR:varied} and~\ref{fig:p:Lgtr5H:xUV:xIR:varied:thermal}. 
However, for UV and IR cuts shown in Figs.~\ref{fig:p:Lgtr5H:xUV:xIR:varied} and~\ref{fig:p:Lgtr5H:xUV:xIR:varied:thermal}, the variation of the $p$ values is still visible. This seems to indicate that the string power spectrum does not follow an ideal power law for the intervals chosen in Figs.~\ref{fig:p:Lgtr5H:xUV:xIR:varied} and~\ref{fig:p:Lgtr5H:xUV:xIR:varied:thermal}. 
Nevertheless, the fit results for two benchmark scenarios, assuming the linear-log hypothesis similarly to the case of the axion, for varying UV cuts are presented in Table~\ref{tab:p:benchmark:fit:4096}. One notes that the fit was done in the same time interval as in~\cite{Kim:2024wku}, and the logarithmic growth is more pronounced in the benchmark scenario using thermal pre-evolution.

\begin{table}[tbh]
\centering
  \renewcommand{\arraystretch}{1.15}
      \addtolength{\tabcolsep}{0.35pt} 
\scalebox{0.90}{
\begin{tabular}{l|c|c}  
\hline
Pre-evolution type & $(x^\text{str}_\text{IR},\, x^\text{str}_\text{UV})$ & Linar-log fit for strings of $H\ell \geq 5$
\\[5pt]
(see Figs.~\ref{fig:p:Lgtr5H:xUV:xIR:varied} and~\ref{fig:p:Lgtr5H:xUV:xIR:varied:thermal}) &  & in the interval, $\log\frac{m_r}{H} = [6.5,\,  7.8] $ 
\\[5pt] 
\hline\hline
             		& (15,\, 16)  &  $p \sim (1.507 \pm 0.077) + (0.131 \pm 0.011) \log\frac{m_r}{H}$ 
\\[5pt]
    Fat-string	& (15,\, 20)  &  $p \sim (1.387 \pm 0.097) + (0.139 \pm  0.013) \log\frac{m_r}{H}$ 
\\[5pt]
			& (15,\, 24)  &  $p \sim (1.174 \pm 0.124) + (0.161 \pm 0.017) \log\frac{m_r}{H}$ 
\\[8pt]
\hline
      			& (15,\, 16)  & $p \sim (0.766 \pm 0.074) + (0.226 \pm 0.001) \log\frac{m_r}{H}$
\\[5pt]
    Thermal 	& (15,\, 20)  & $p \sim (0.536 \pm 0.094) + (0.249 \pm 0.013) \log\frac{m_r}{H}$
\\[5pt]
			&  (15,\, 24) & $p \sim (0.379 \pm 0.120) + (0.262 \pm 0.016) \log\frac{m_r}{H}$ 
\\\hline
\end{tabular}
}
\caption{\small The fit result taking the linear-log hypothesis for two benchmark simulations on $N^3 = 4096^3$ differing by the relaxation schemes. The presented errors  merely represent the quality of the fit for the given choices of cuts on $x_\text{IR}$, $x_\text{UV}$ and the fitting intervals.
}
\label{tab:p:benchmark:fit:4096} 
\end{table}

\begin{figure}[tp]
\begin{center}
\includegraphics[width=0.48\textwidth]{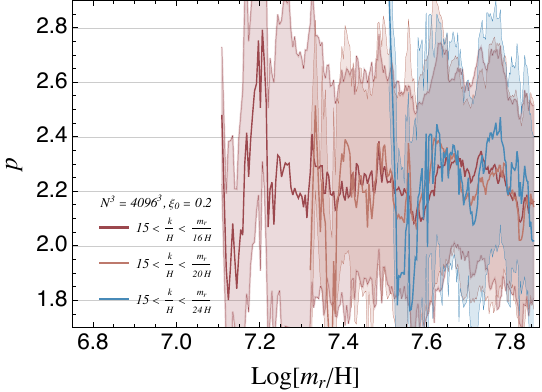}\quad
\includegraphics[width=0.48\textwidth]{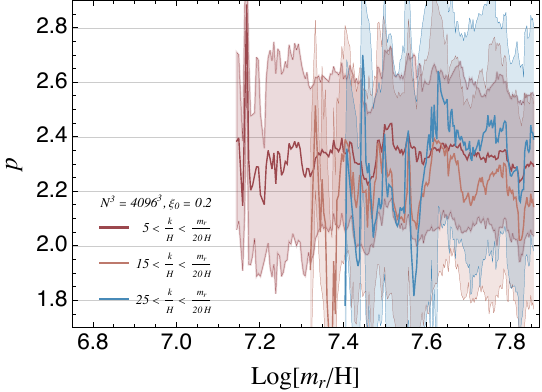}
\caption{\small The averaged spectral index $p$ of the string power spectrum for sub-Hubble strings of $\ell < H^{-1}$ obtained by fitting the data within the momentum interval, $x^\text{str}_\text{IR} < k/H < m_r/(x^\text{str}_\text{UV}H)$. In the left (right) panel, $x^\text{str}_\text{IR}$ ($x^\text{str}_\text{UV}$) is varied while $x^\text{str}_\text{UV}$ ($x^\text{str}_\text{IR}$) is fixed. Both panels were based on data assisted by the fat-string pre-evolution for the benchmark scenario of $\xi_0 = 0.2$.
}
\label{fig:p:Lless1H:xUV:xIR:varied}
\end{center}
\end{figure}

A similar exercise for sub-Hubble strings of $\ell < H^{-1}$ was done with $\ell_\text{min}= 0$ and $\ell_\text{max}= H^{-1}$ and the result is illustrated in Fig.~\ref{fig:p:Lless1H:xUV:xIR:varied}. Due to small values of $H\ell \lesssim 1$, the minimum number of data points for the fit is available only at late times. As is evident in Fig.~\ref{fig:p:Lless1H:xUV:xIR:varied}, looking at only central values, the scaling behaviors do not look same as those for long strings although it is difficult to draw a definite conclusion due to large statistical errors. Recall that the fraction of sub-Hubble strings $\xi_\ell/\xi$ is small, as was shown in Fig.~\ref{fig:loopstrings:fat:4096}.

\section{From strings to axions}
\label{sec:strings:axions}

%
The string spectrum that we have obtained solely from lattice simulations will have to be related to the axion spectrum. The simulation-driven analyses of strings and axions~\cite{Kim:2024wku} indicate that two spectral indices of power law profiles are roughly related by $q \sim p - \Delta$ at late times where $\Delta$ takes an order one value which apparently depends on the choice of UV, IR cuts. 
In this section, we present our analytic derivation for such a relation. 

With three assumptions, namely the absence of the expansion of the universe, small amplitudes compared to the wavelength of a string~\cite{Battye:1993jv}, and the Kalb-Ramond (KR) type interaction between axions and strings~\cite{Kalb:1974yc,Davis:1988rw}, an analytic calculation for predicting the axion spectrum is plausible. The string oscillations of $k \gg H$ will not be affected by the expansion of the universe (we numerically checked for a sinusoidal oscillation of a single frequency by comparing two simulation results with and without the Hubble expansion), whereas the oscillations of $k \lesssim H$ will likely be frozen in the expanding universe due to the Hubble expansion.
The second assumption on small amplitudes with respect to the wavelength is empirically seen to be well satisfied in the region for $k/H \gg 1$, as is indicated in Fig.~\ref{fig:strPS:4096}, and it allows us to approximate strings with almost straight ones (see also Section~\ref{app:sec:singlestring} for the related discussion).
The KR action should be a good approximation in the thin string limit, namely the string width of the order of $m_r^{-1}$ should be much smaller than the radius of the curvature of the fluctuation. While the thin string limit should be easily met for fluctuations of the order of the Hubble length or larger, for instance, $m_r^{-1}/H^{-1} \sim e^{-\log \frac{m_r}{H}} \lesssim 10^{-3}$ at late times of $\log\frac{m_r}{H} \gtrsim 6$, it may be less trivially satisfied for fluctuations of much shorter wavelengths than the Hubble length.

We emphasize that our analytic derivation is based on three assumptions and, as a result, the description becomes invalid beyond the phase space that is consistent with those assumptions. Nevertheless, it should provide a strong qualitative support for the correlation between power spectra of strings and axions. 

\subsection{Linking strings to axion spectrum}
\label{app:sec:strings:to:axions}

Having our assumptions outlined above, we provide a detailed derivation for the analytic relation between the string spectral index $p$ and the axion spectral index $q$. Our analytic result should be considered as the leading order contribution in terms of the expansion parameters parametrizing the validity of assumptions. The non-negligible effects from the regime where some of the above three assumptions are violated will be discussed qualitatively.

\subsubsection{Almost straight strings}

We first derive the form of the Fourier amplitude, $\vec{T}_n^\ell$ in Eq.~\ref{eq:T:Fourieramp}, of a generic static string in an almost straight string limit. When the amplitudes of fluctuations of strings are much smaller than the string wavelength, the assumption for an almost straight string with sinusoidal fluctuations in transverse directions can be justified. 
What it meant by `static' is that the time dependence of the fluctuation is not considered.

Imagine an almost straight string aligned along the $z$-axis. The configuration of such a string can be parametrized by a vector as a function of $z$, $\vec{X} (z) = ( \vec{X}_\perp (z),\, z )$ where the transverse motion can be written in the Fourier form,
\begin{equation}\label{app:eq:X:straight:static}
   \vec{X}_\perp (z) = \sum_{n=-\infty}^{\infty} \frac{1}{n}\vec{A}_n e^{i \frac{2\pi n}{L} z}~,
\end{equation}
where $L$ is the periodicity in the $z$-direction, or the wavelength of the lowest mode, and $\vec{A}_n = (A^x_n, \, A^y_n)$, $\vec{A}_0 = 0$. 
The fact that the fluctuation in Eq.~\ref{app:eq:X:straight:static} is real implies $\vec{A}_{-n} = - \vec{A}_n^*$.
The single parameter $s$ parametrizing the path along the generic string is related to $z$ as
\begin{equation}
  \frac{ds}{dz} = \left | \frac{d\vec{X}}{dz} \right | = \left ( 1- (2\pi)^2 \sum_{n,n'=-\infty}^\infty \frac{\vec{A}_n \cdot \vec{A}_{n'}}{L^2} e^{\frac{2\pi i(n+n')z}{L}} \right )^{1/2}~.
\end{equation}
Integration $\int_0^z dz'(ds/dz')$ gives rise to
\begin{equation}
 s = \frac{\ell}{L} z - \frac{(2\pi)^2}{2} \sum_{m=1}^{\infty}\frac{L}{m\pi} \sin \left (\frac{\pi m z}{L} \right ) 
 \sum_{n = -\infty}^\infty \left ( e^{\frac{i \pi m z}{L}} \frac{\vec{A}_n\cdot \vec{A}_{m-n}}{L^2} + e^{-\frac{i\pi m z}{L}} \frac{\vec{A}^*_n \cdot \vec{A}^*_{m-n}}{L^2}  \right ) + \cdots~,
\end{equation}
where $\cdots$ denotes higher order terms in $\epsilon \equiv A/L$, the ratio of the amplitude to the wavelength of the lowest mode.
The Fourier amplitude in Eq.~\ref{eq:T:Fourieramp} of an almost straight string is estimated to be
\begin{equation}\label{app:eq:Tn:straight}
\begin{split}
 \vec{T}^\ell_{n,\perp} &= \frac{2\pi i}{\ell} \Big [ \vec{A}_n + n L \frac{(2\pi)^2}{2} \sum_{m=1}^\infty \frac{L}{m\ell} 
 \sum_{q=-\infty}^\infty \Big \{ \frac{\vec{A}_{n-m} - \vec{A}_n}{L} \frac{\vec{A}_q \cdot \vec{A}_{m-q}}{L^2}
 \\[5pt]
 &\hspace{7cm} - \frac{\vec{A}_{n+m} - \vec{A}_n}{L} \frac{\vec{A}^*_q \cdot \vec{A}^*_{m-q}}{L^2} \Big \} + \cdots \Big ],
 \\[5pt]
 \vec{T}^\ell_{n, z} &= \frac{L}{\ell} \Big [ \delta_{0n} + \left ( 1- \delta_{0n} \right ) \frac{L}{\ell} \frac{(2\pi)^2}{2}
 \sum_{q=-\infty}^\infty \frac{\vec{A}_q \cdot \vec{A}_{n-q}}{L^2} + \cdots \Big ]~,
\end{split}  
\end{equation}
where $\cdots$ denotes higher order terms in $\epsilon$.

\subsubsection{Axion spectrum from almost straight strings}
\label{app:sec:predicted:axion}

Axions radiated from almost straight cosmic strings can be analytically studied, provided the interaction between axions and cosmic strings. An almost straight string along the $z$-axis, oscillating in $xy$-plane (perpendicular to $z$-axis), in the non-expanding universe can be written as~\cite{Battye:1993jv,Vilenkin:1986ku,Sakellariadou:1991sd}
\begin{equation}\label{app:eq:X:straight:dynamic}
 \vec{X}_\perp (\tau,\, \sigma) = \frac{1}{2} \sum_{n=-\infty}^\infty \frac{1}{n\Omega} 
 \left ( \vec{a}_n e^{in\Omega (\sigma-\tau)} + \vec{b}_n e^{in\Omega (\sigma+\tau)} \right )~,
\end{equation}
where $\tau,\, \sigma$ are the world-sheet coordinates of a string and $\Omega = 2\pi/L$, $\vec{a}_n = (a_n^x,\, a_n^y)$ with $\vec{a}_0=0$ (similarly for $\vec{b}_n$). Since the expansion of the universe is ignored, $\tau$ is the same as the cosmic time.
Note that unlike the static string in Eq.~\ref{app:eq:X:straight:static}, the oscillations were decomposed into those of left- and right-movers in the dynamic parametrization in Eq.~\ref{app:eq:X:straight:dynamic}. 

Assuming the KR type interaction, the radiation power per length into axions of frequency $\omega_n = \frac{2\pi n}{L}$ from a string of the fixed length $L$ is given by at leading order~\cite{Battye:1993jv,Vilenkin:1986ku,Sakellariadou:1991sd},
\begin{equation}\label{app:eq:Pn}
 \frac{dP_{L,\, n}}{dz} = \frac{\pi^3 f_a^2}{2L} n \sum_{\substack{|m|<n,\\m+n\, \text{even}}}
 \left ( |\vec{a}_{\frac{m+n}{2}}|^2  |\vec{b}_{\frac{m-n}{2}}|^2 -  |\vec{a}_{\frac{m+n}{2}} \cdot \vec{b}_{\frac{m-n}{2}}|^2 +  |\vec{a}^*_{\frac{m+n}{2}}\cdot \vec{b}_{\frac{m-n}{2}}|^2 \right )~.
\end{equation}
The radiation power at leading order  in Eq.~\ref{app:eq:Pn} is proportional to only cross terms between left- and right-movers of frequencies $\omega_{(n+m)/2}$ and $\omega_{(m-n)/2}$, respectively.
This implies that axions can be thought of being produced through the annihilations of the left- and right-movers of all possible frequencies. 
The power spectrum can be obtained by summing over all frequencies, 
\begin{equation}
  P_L (k) = L \sum_{n=1}^\infty \frac{dP_{L,\, n}}{dz} \delta \left ( k - \frac{2\pi n}{L} \right )~,
\end{equation}
where the delta function was to express in terms of the momentum $k$ and the string length $\ell$ is equal to $L$ at leading order in $\epsilon$.
The total radiation power of the momentum $k$ from $N$ strings of various lengths in the string network is given by
\begin{equation}\label{app:eq:power:a:4pt}
\begin{split}
  \langle P(k) \rangle &= \int_0^\infty d\ell\, N  \rho(\ell) \langle P_\ell(k) \rangle
  \\[5pt]
  &= N \sum_{n=1}^\infty \frac{\pi^3 f_a^2}{2} n \int_0^\infty d\ell\, \rho(\ell) \delta \left ( k - \frac{2\pi n}{\ell} \right )
  \\[5pt]
  &\hspace{2cm} \times \sum_{\substack{|m|<n,\\m+n\, \text{even}}}
 \langle |\vec{a}_{\frac{m+n}{2}}|^2  |\vec{b}_{\frac{m-n}{2}}|^2 -  |\vec{a}_{\frac{m+n}{2}} \cdot \vec{b}_{\frac{m-n}{2}}|^2 +  |\vec{a}^*_{\frac{m+n}{2}}\cdot \vec{b}_{\frac{m-n}{2}}|^2 \rangle~,
\end{split}
\end{equation}
where $\langle \cdots \rangle$ denotes the average done over many simulations of the string network and $\rho(\ell)$ is the probability density of a string of a length $\ell$. 
Using the invariance under the string parametrization with constant shiftings of $\sigma$ and $\tau$, inversions of them ($\sigma \rightarrow -\sigma$ and $\tau \rightarrow -\tau$) and the rotation about $z$-axis along with the realness of $\vec{X}_\perp$, the total radiation power spectrum can be written as
\begin{equation}\label{app:eq:power:a:4pt:cn}
\begin{split}
  \langle P(k) \rangle &= N \sum_{n=1}^\infty 2\pi^3 f_a^2 n \int_0^\infty d\ell\, \rho(\ell) \delta \left ( k - \frac{2\pi n}{\ell} \right )
 \sum_{\substack{|m|<n,\\m+n\, \text{even}}} c_{\frac{m+n}{2}}(\ell) c_{\frac{m-n}{2}}(\ell)~,
\end{split}
\end{equation}
where $\langle a^i_n a^j_m \rangle = \langle b^i_n b^j_m \rangle = - c_n(\ell)\delta^{ij}\delta_{(n+m)0}$ ($c_n$ is real). In the derivation of Eq.~\ref{app:eq:power:a:4pt:cn} from Eq.~\ref{app:eq:power:a:4pt}, $\vec{a}_n$ and $\vec{b}_n$ were assumed to follow the Gaussian distributions for $n\neq 0$ which allow the decomposition of 4-point correlations among them in terms of 2-point correlations (known as Wick theorem). A significant deviation from this assumption may invalidate our derivation although we believe that this is not the case~\footnote{We have numerically checked the distribution of the Fourier amplitude $\vec{T}^\ell_{n,\perp}$ (denoted by $T$ below) as a proxy of $\vec{A}_n$ in Eq.~\ref{app:eq:X:straight:static}. It is observed to follow the Gaussian distribution well for the momenta of interest (the ratio of moments $\langle T^4 \rangle/\langle T^2 \rangle^2 \sim 3$, which is the value for the Gaussian, was also numerically checked).}.
On the other hand, plugging $\vec{A}_n$, obtained by re-writing $\vec{X}_\perp$ in Eq.~\ref{app:eq:X:straight:dynamic} in the form of Eq.~\ref{app:eq:X:straight:static}, into Eq.~\ref{app:eq:Tn:straight}, the Fourier amplitudes are approximated as
\begin{equation}
\begin{split}
 \vec{T}^\ell_{n,\perp} &=\frac{2\pi i}{\ell} \frac{L}{4\pi} \left ( \vec{a}_n e^{-\frac{2\pi i n t}{L}} + \vec{b}_n e^{\frac{2\pi i n t}{L}}  \right ) + \cdots~,
 \\[5pt]
 \vec{T}^\ell_{n,z} &= \frac{L}{\ell} \delta_{n0} + \left ( \frac{L}{\ell} \right )^2 \frac{1}{8} \left ( 1 - \delta_{0n} \right )
 \sum_{q=-\infty}^\infty \left ( \vec{a}_q\cdot\vec{b}_{n-q} e^{\frac{2\pi i (n-2q)t}{L}}
 + \vec{a}_{n-q}\cdot \vec{b}_q e^{-\frac{2\pi i (n-2q)t}{L}} \right ) + \cdots~,
\end{split}
\end{equation}
where $\tau = t$, $\sigma = z + \cdots$, and $\ell = L + \cdots$ at leading order in $\epsilon$.
The string spectrum $\mathcal{T}^\ell_n = \langle |\vec{T}_n^\ell |^2 \rangle$ is approximated, at leading order in $\epsilon$, 
\begin{equation}\label{app:eq:T:ab}
  \mathcal{T}^\ell_n = \frac{1}{4} \left ( \langle |\vec{a}_n|^2 \rangle + \langle |\vec{b}_n|^2 \rangle \right ) + \cdots~
  = c_n(\ell) + \cdots~\text{for}~ n \neq 0~.
\end{equation}
Therefore, the axion power spectrum can be written as, using the expression in Eq.~\ref{app:eq:T:ab},
\begin{equation}
\begin{split}
  \langle P(k) \rangle &= N \sum_{n=1}^\infty 2\pi^3 f_a^2 n \int_0^\infty d\ell\, \rho(\ell) \delta \left ( k - \frac{2\pi n}{\ell} \right )
  \sum_{\substack{|m|<n,\\m+n\, \text{even}}} \mathcal{T}^\ell_{\frac{n+m}{2}}  \mathcal{T}^\ell_{\frac{n-m}{2}}~.
\end{split}
\end{equation}
The axion power spectrum from an almost straight string for $k \gg H$ can be estimated by dividing the total radiation power by the volume of the space, namely $\Gamma_a^\text{str.} (k) = \langle P(k) \rangle/\rm{Vol.}$,
\begin{equation}\label{app:eq:Gamma:axion:T}
  \frac{\Gamma_a^\text{str.} (k)}{m_r^2 f_a^2} = 
  8\pi^3 \xi \left ( \frac{m_r}{H} \right )^{-2} \int_0^\infty \frac{d(H\ell)}{H\ell_\text{ave.}} \frac{\rho(\ell)}{H}
  \sum_{n=1}^\infty n H \delta \left ( k - \frac{2\pi n}{\ell} \right ) 
  \sum_{\substack{|m|<n,\\m+n\, \text{even}}} \mathcal{T}^\ell_{\frac{n+m}{2}}  \mathcal{T}^\ell_{\frac{n-m}{2}}~,
\end{equation}
where $\xi = \ell_\text{tot}t^2/\text{Vol.}$, $\ell_\text{tot} = N \ell_\text{ave.}$ and $H=1/(2t)$ were used.

\begin{figure}[tp]
\begin{center}
\includegraphics[width=0.48\textwidth]{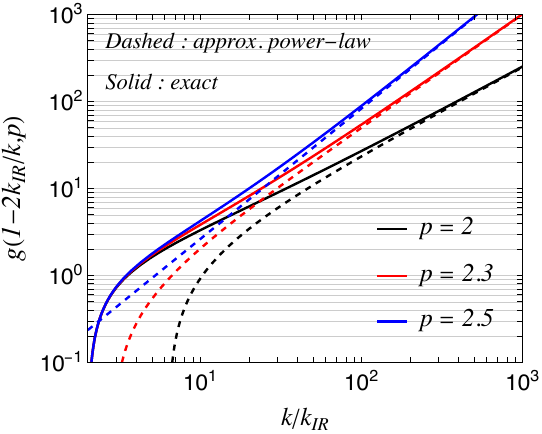} \quad
\includegraphics[width=0.48\textwidth]{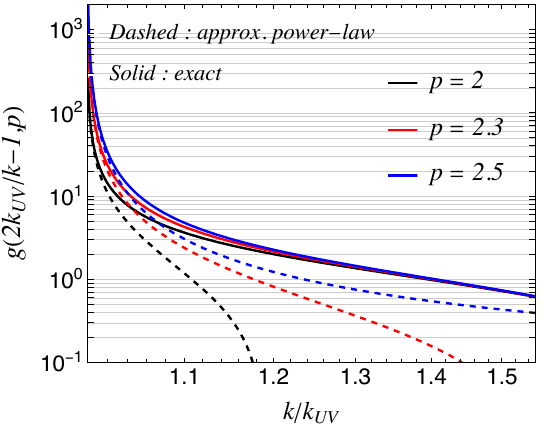}
\caption{\small The exact hypergeometric function $g(1-2k_\text{IR}/k,\, p)$ (left) in Eq.~\ref{app:eq:Gamma:a:threecases} for various $p$ values (solid lines) and their leading scaling approximations when $k_\text{IR}/k \ll 1$ (dashed). Similarly for $g(2k_\text{UV}/k - 1,\, p)$ (right)
}
\label{fig:g:approx}
\end{center}
\end{figure}

If we assume that the power law scaling of the axion spectrum is originated from that of the string spectrum, we may parametrize the power spectrum $\mathcal{T}_n^\ell$ as
\begin{equation}\label{app:eq:Tn:powerlaw}
\begin{split}
\mathcal{T}_n^\ell = &
\left\{ 
\begin{array} {lll}
 B_\ell \left ( \displaystyle\frac{k_n}{H} \right )^{-p} & \text{for} & k_\text{IR} \leq k_n \leq k_\text{UV}
\\[15pt]
\hspace{0.14in}  0 & & \text{otherwise}~,
\end{array}
\right.
\end{split}
\end{equation}
where $B_\ell$ is a constant. The form of $\mathcal{T}_n^\ell$ in Eq.~\ref{app:eq:Tn:powerlaw} leads to the axion power spectrum,
\begin{equation}\label{app:eq:axion:powerlaw}
\begin{split}
  \frac{\Gamma_a^\text{str.} (k)}{m_r^2 f_a^2} =&\ 
  8\pi^3 \xi \left ( \frac{m_r}{H} \right )^{-2} \int_0^\infty \frac{d(H\ell)}{H\ell_\text{ave.}} \frac{\rho(\ell)}{H}
  B_\ell^2 \left ( \frac{2\pi}{H \ell} \right )^{-2p}
  \sum_{n=1}^\infty n H \delta \left ( k - \frac{2\pi n}{\ell} \right ) 
  \\[5pt]
  & \times
  \sum_{{m+n\, \text{even}}}
  \left ( \frac{n+m}{2} \right )^{-p} \left ( \frac{n-m}{2} \right )^{-p}~,
\end{split}
\end{equation}
where the integer $m$ is restricted to $|m| \leq \text{min} (\ell k_\text{UV}/\pi -n,\, n- \ell k_\text{IR}/\pi)$ due to the cuts on the interval in Eq.~\ref{app:eq:Tn:powerlaw}. To extract the power law scaling in the momentum, $k_n = \frac{2\pi n}{\ell}$, we can focus on only the $n$ dependences in the continuum limit which is
\begin{equation}
\begin{split}
  \Gamma_a^\text{str.} (k) &\propto 
  n^{2-2p} \int_{-\infty}^\infty \frac{dm}{n} \left (1- \frac{m^2}{n^2} \right )^{-p}~,
\end{split}
\end{equation}
where $m$ is constrained as was mentioned above, or $|m|/n \leq \text{min} ( 2 k_\text{UV}/k -1,\, 1- 2 k_\text{IR}/k )$. 
The integral is symmetric in $m$. It can be re-written as, depending on the upper bound on $m$,
\begin{equation}
\begin{split}\label{app:eq:Gamma:a:threecases}
  \Gamma_a^\text{str.} (k) &\propto  k^{2-2p}
\left\{ 
\begin{array} {lll} 
 g \left ( 1-\displaystyle\frac{2k_\text{IR}}{k},\, p \right ) & \text{for} & 2 k_\text{IR} \leq k \leq k_\text{IR} + k_\text{UV}
\\[15pt]
 g \left ( \displaystyle\frac{2k_\text{UV}}{k} - 1,\, p \right ) & \text{for} & k_\text{IR} + k_\text{UV} \leq k \leq 2 k_\text{UV}
\\[15pt]
\hspace{0.14in}  0 & & \text{otherwise}~,
\end{array}
\right.
\end{split}
\end{equation}
where $g(x,\, p) = 2 \int_0^x du (1- u^2 )^{-p}$ which gives a hypergeometric function. Note that cutoffs on the string spectrum, $k_\text{IR} < k < k_\text{UV}$, not only translate to shifted cutoffs, $2k_\text{IR} < k < 2k_\text{UV}$, in the axion spectrum, but also the parametric behavior of $\Gamma_a^\text{str}$ changes around  $k_\text{IR}+k_\text{UV}$. 
Assuming large separations of $k$ from both UV and IR cutoffs, 
or $k_\text{IR} \ll k \ll k_\text{UV}$, the function $g(x,\, p)$ in the first line of Eq.~\ref{app:eq:Gamma:a:threecases} can be expanded around $x=1$, namely
\begin{equation}
g(x,\, p) = \frac{\pi^{3/2} \csc(p\pi)}{\Gamma(\frac{3}{2}-p) \Gamma(p)} - \frac{2^{1-p}}{(1-x)^{p-1}}
\sum_{n=0}^{\infty} \frac{\Gamma(p+n)}{\Gamma(p)} \frac{(1-x)^n}{n! (n+1-p)2^n}~,
\end{equation}
where the leading $k$-dependence of the function $g$ appears as $g(x,\, p) \propto (1-x)^{1-p}$ which leads to 
$g(1-2k_\text{IR}/k,\, p) \propto k^{p-1}$. We find that the exact evaluation of $g$ can be well approximated by a scaling $\sim k^{p-1}$, for instance, as long as $k \gtrsim 10 k_\text{IR}$ for $p \gtrsim 2$ (which is the range of interest) as is illustrated in Fig.~\ref{fig:g:approx}. This minimal effective IR cutoff (that ensures the simple scaling) is a bit higher than $2k_\text{IR}$ in Eq.~\ref{app:eq:Gamma:a:threecases} and it varies depending on the value of $p$. This will be translated to $5\times(2k_\text{IR}) \leq x_\text{IR} H$ in terms of notations in~\cite{Kim:2024wku}~\footnote{It is an inequality since $x_\text{IR}$ was our arbitrary choice in~\cite{Kim:2024wku} whereas  $5\times(2k_\text{IR})$ here is a minimal IR cutoff, for a given $p$, for having a simple scaling.}. 
The second term in Eq.~\ref{app:eq:Gamma:a:threecases} contributes to the UV region $k_\text{UV} \lesssim k < 2k_\text{UV}$. 
The right panel of Fig.~\ref{fig:g:approx} shows that the function $g$ does not have a simple scaling behavior above $k_\text{UV}$ and thus $k_\text{UV}$ sets the maximum effective UV cutoff for ensuring a simple scaling.
Therefore, we expect that the axion spectrum from cosmic strings scales as
\begin{equation}\label{app:eq:power:strings:to:axion}
\begin{split}
  \Gamma_a^\text{str.} (k) &\propto  k^{2-2p} \times k^{p-1} = k^{-(p-1)}~,
\end{split}
\end{equation}
in the interval, $\alpha \times (2k_\text{IR}) \lesssim k \lesssim k_\text{UV}$ ($\alpha$ as a value of several depending on $p$) under assumptions that we made for the analytic derivation. The effective IR cutoff $\alpha \times (2k_\text{IR})$ seems to suggest a higher IR cutoff in the axion spectrum by the factor of $2\alpha$. The fact that our derivation is valid for $k\gg H$ is consistent only with an aggressive IR cutoff $x_\text{IR}$ in the axion spectrum. 
On contrary, a mild value of IR cutoff $x_\text{IR}$ in the axion spectrum is associated with a too low IR cutoff of the string spectrum where our assumption may not be valid anymore. For instance, $x_\text{IR} = 15$, which was taken in~\cite{Kim:2024wku} as one of choices for IR cutoff, will be consistent with $k_\text{IR} \lesssim 2H$ (for $\alpha \sim 5$). In this situation, the comparison with the result of the axion spectrum in literature should be taken with a grain of salt.
%
%

\subsection{Summary}

The relation between two spectral indices $q = p - 1$ in Eq.~\ref{app:eq:power:strings:to:axion} is a consequence of a few factors including the peculiar axion radiation mechanism from oscillating strings. As was described in Section~\ref{app:sec:strings:to:axions}, the oscillations on an almost straight string are superposition of two types modes for the left- and right-movers. The radiation power into the axion of the frequency of $\omega_n = 2\pi n/L$ is obtained by summing over all the cross terms between the left- and right-movers of frequencies $\omega_{(n + m)/2}$ and $\omega_{(n-m)/2}$, respectively. It can be viewed that the axions are produced through the annihilations of the left- and right movers. Keeping only relevant parts in Eq.~\ref{app:eq:axion:powerlaw}, it takes the form for a large $n$ (equivalently, a large momentum), 
\begin{equation}\label{eq:powercounting:Gamma:a}
  \Gamma_a^\text{str.}(n) \propto n \sum_{m=1}^{n-1} \frac{1}{(n+m)^p (n-m)^p} \sim n^{-(p-1)}~,
\end{equation}
where two terms in the summation are originated from the product of two string power spectra scaling like $\mathcal{T}^\ell \sim k^{-p}$.
The larger amplitude at the frequency of $\omega_{(n-m)/2}$ is canceled by the smaller amplitude at the higher frequency of $\omega_{(n+m)/2}$, and summing over all those cross terms, along with an overall $k_n \propto n$ factor in $\Gamma_a^\text{str.}$, leads to the resulting axion power spectrum in a nontrivial way.

However, the schematic form in Eq.~\ref{eq:powercounting:Gamma:a} indicates a caveat in our analytic calculation. While our analytic calculation was strictly restricted to the high momentum region $k\gg H$ of the string power spectrum, the form in Eq.~\ref{eq:powercounting:Gamma:a} shows that the contribution to the axion power spectrum at a high momentum can also get affected by the low-momentum region of the string spectrum where our assumptions on the absence of the Hubble expansion and the small ratio of the amplitudes to the wavelength can be violated (see also Appendix~\ref{app:sec:singlestring} for the related discussion). The aforementioned shifted interval $\alpha \times (2k_\text{IR}) \lesssim k \lesssim k_\text{UV}$ is the direct consequence of this property. 
A moderate value of the cutoff $x_\text{IR}$ in the axion spectrum is consistent with a value of momentum cutoff as low as $k_\text{IR} \sim H \times x_\text{IR}/(2\alpha)$.
In reality, there exists non-vanishing string spectrum in the low momentum region where an almost straight string may not be valid anymore, and additionally, a correction to the KR approximation in the high momentum region may exist as well although addressing those is beyond the scope of this work.
In a more general situation, the predicted spectral index $q$ of the axion power spectrum may take a form,
\begin{equation}\label{eq:q:p}
   q = p - 1 - \Delta^\text{str.}_\text{nonpert.}~,
\end{equation}
where $\Delta^\text{str.}_\text{nonpert.}$ accounts for the correction from the non-perturbative phase space which is not captured in our approximation.
Comparing results in Table~\ref{tab:p:benchmark:fit:4096} and those in~\cite{Kim:2024wku}, without carefully coordinating proper fit intervals, $\Delta^\text{str.}_\text{nonpert.} \sim \text{a few}\times 0.1$ looks a ballpark value.

\section{Conclusion}

The logarithmic growth of the spectral index of the axion power spectrum was an intriguing observation that was made in recent lattice simulations. The confirmation of it will have a significant impact on the cosmology of the QCD axion dark matter. 
In this work, we have presented a new scaling observable, the spectral index of the string power spectrum, which exhibits a similar logarithmic growth as our main result of this work. Since axions are radiated from the cosmic strings, it is naturally expected that the characteristic property of the axion spectrum is closely related to that of strings. In order to make such a relation more concrete, we have derived the analytic relation between two power spectra of strings and axions under some assumptions and discussed the validity of those.
Improving the analytic relation will be interesting and relevant for better understanding the recently observed logarithmic growth in the axion spectrum.

In this work, we have initiated the statistical analysis of the causal dynamics of the cosmic strings. 
While it will be interesting to look for as many statistical distributions as possible which can reveal the UV nature of cosmic strings or the microscopic nature of their dynamics, we have made a few basic demonstrations for singe string motions.
The fragmentation rate of a string loop into smaller loops as a function of the string curvature and the distributions of the branched loop sizes may be useful to probe the UV nature of cosmic strings and to understand the string intercommutation itself. We leave the detailed study of it to the future work.

\section*{Acknowledgments}
HK and MS were supported by National Research Foundation of Korea (NRF) under Grant Number RS-2024-00450835.
The simulation in this work was supported by KISTI National Supercomuting Center under Project Number KSC-2024-CRE-0278.

\appendix

\section{More on simulation setup}
\label{app:setups}
The simulation setup adopted in our study is identical to those in~\cite{Kim:2024wku} where more simulation details can be found.

\subsection{Equation of motion}
The equation of motion in Eq.~\ref{eq:Lag:orig} can be more conveniently implemented in the numerical simulation by rescaling the field and spacetime coordinates as $\phi \rightarrow f_a \phi$, $t \rightarrow m_r^{-1}t$, and $\vec{x} \rightarrow m_r^{-1} \vec{x}$,
\begin{equation}
 \ddot{\phi} + 3 H \phi - \frac{1}{R^2} \nabla^2 \phi + \phi \left ( |\phi|^2 - \frac{1}{2} \right ) = 0~,
\end{equation}
where dot is the differentiation with respect to the dimensionless time $t$ and the gradient $\nabla$ is with dimensionless comoving coordinates. 
The simulation is more efficient with the equation of motion in terms of $\psi(\tau, \vec{x}) = R \phi(t,\vec{x})$,
\begin{equation}
 \psi'' - \nabla^2 \psi + \psi \left ( |\psi|^2 - \frac{R^2}{2} - \frac{R''}{R} \right ) = 0~,
\end{equation}
where the last term $R''/R$ vanishes in the radiation-dominated era, since the Courant-Friedrichs-Lewy (CFL) condition in this situation is relaxed to $\Delta \tau \lesssim \Delta x$, as opposed to $\Delta t \lesssim R \Delta x$ in the equation in terms of $\phi$. In the latter, the CFL condition requires a small time step size in the early time of the simulation due to a small scale factor.

\subsection{Fat string pre-evolution}
The string network formed around the time, when the random field configuration was generated, contains a high level contamination from noisy short distance structures. Cleaning up or relaxing those contaminations is a necessary step, called a pre-evolution, to make the approach of the string network toward the scaling solution more efficient. In the fat-string approach that we adopt in our simulation~\cite{Gorghetto:2020qws}, the string core size $m_r^{-1}$ during the pre-evolution is engineered to scale as $R(t) \propto t \sim H^{-1}$ so that the ratio $\frac{m_r}{H}$ remains constant. Since all three quantities $m_r^{-1}$, $R$, and $H^{-1}$ scale the same in time $t$ in the fat-string approach, the pre-evolution can run as long as it can to reduce the contamination. The outcome of the pre-evolution serves the initial condition for the physical string evolution in the radiation-dominated era. In our simulation, the pre-evolution is stopped when the requested number of strings per Hubble patch $\xi_0$ is reached. 
We choose $\xi_0 = 0.2$ as our benchmark scenario, based on the study in~\cite{Kim:2024wku}.
The random field (and its velocity) configuration is generated with the momentum cutoff $k_\text{max}= m_r$ as in~\cite{Kim:2024wku}.

\subsection{Thermal pre-evolution}
In the thermal pre-evolution, the spontaneous breaking of the PQ symmetry is realized by a temperature-dependent potential, $\Delta V = \frac{m_r^2}{6f_a^2} T^2 |\phi|^2$.  After the rescaling of the field and spacetime coordinates along with $T \rightarrow f_a T$, it modifies the equation of motion in terms of $\psi$ as
\begin{equation}
 \psi'' - \nabla^2 \psi + \psi \left ( |\psi|^2 - \frac{R^2}{2} + \frac{1}{6} R^2 T^2 \right ) = 0~,
\end{equation}
where $T^2 = 2\zeta H$ (in terms of rescaled temperature and Hubble parameter) and 
\begin{equation}
   \zeta = \frac{45}{2\pi^2 g_*} \frac{M_p^2 m_r^2}{f_a^4}~,
\end{equation}
where $g_*$ is the relativistic degrees of freedom and $M_p$ is the Planck mass.
In this approach, the correlation length in the string core size unit, $\frac{m_r}{H} |_{T_c}$ at the critical temperature $T_c$ is set by $\zeta$ which we use to choose an appropriate initial condition, 
\begin{equation}
   \left . \frac{m_r}{H} \right |_{T_c} = \frac{2}{3} \zeta~.
\end{equation}
The phase transition time $\tau_c$ is given by $\tau_c^2 = m_r^{-2} \times \frac{4}{3}\zeta$. In our lattice simulation, the initial time was chosen as $\tau_i = 0.1 \tau_c$. 
The random initial values of fields and their velocities were generated assuming the Gaussian random field configuration following the thermal distributions~\cite{Kawasaki:2018bzv}.
The high momentum modes were restricted by the momentum cutoff of $k_\text{max}|_{\tau_i} = 10 m_r$ which is equivalent to $k_\text{max}|_{\tau_c} = m_r$.
We choose $\zeta = 2.12$ as our benchmark scenario, based on the study in~\cite{Kim:2024wku}.

\section{More on string spectrum}
\label{app:sec:morestring}

\begin{figure}[tp]
\begin{center}
\includegraphics[width=0.45\textwidth]{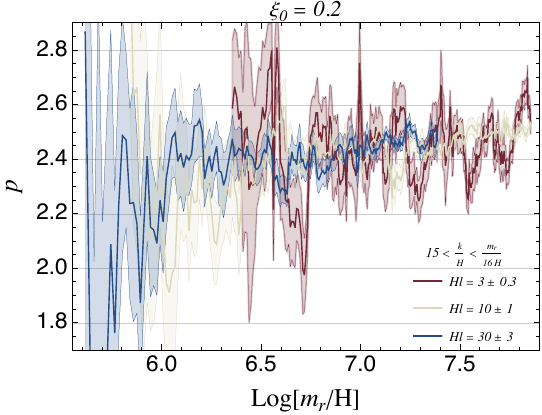} \quad
\includegraphics[width=0.45\textwidth]{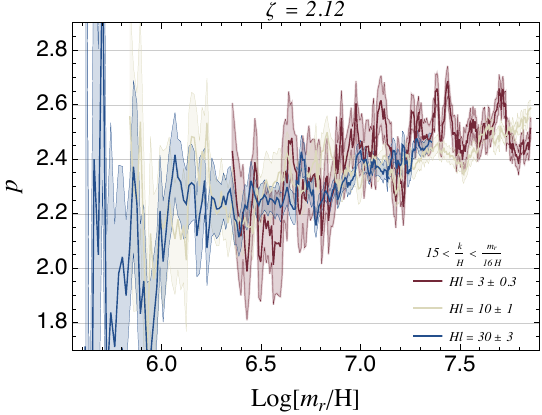}
\\[10pt]
\includegraphics[width=0.45\textwidth]{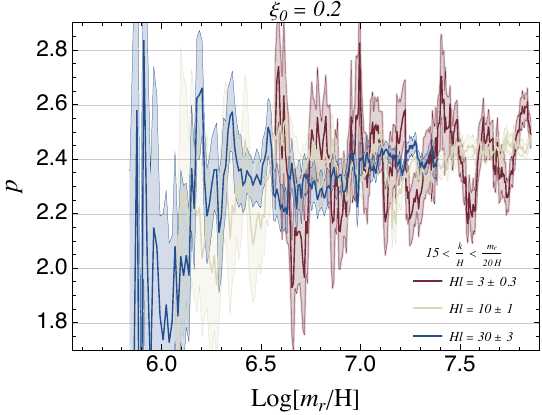} \quad
\includegraphics[width=0.45\textwidth]{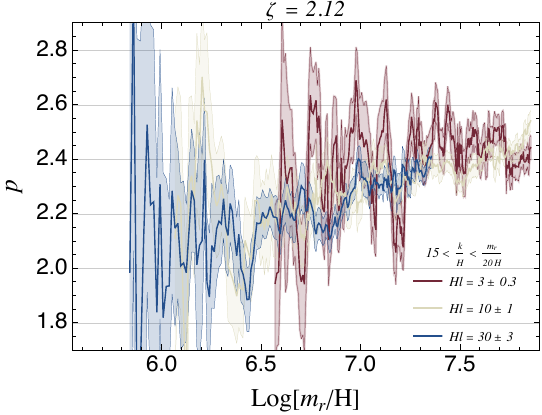}
\\[10pt]
\includegraphics[width=0.45\textwidth]{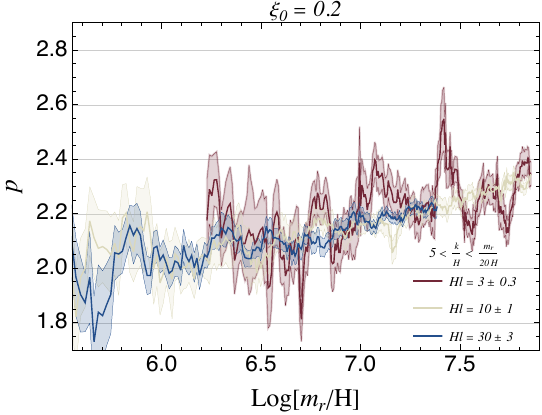} \quad
\includegraphics[width=0.45\textwidth]{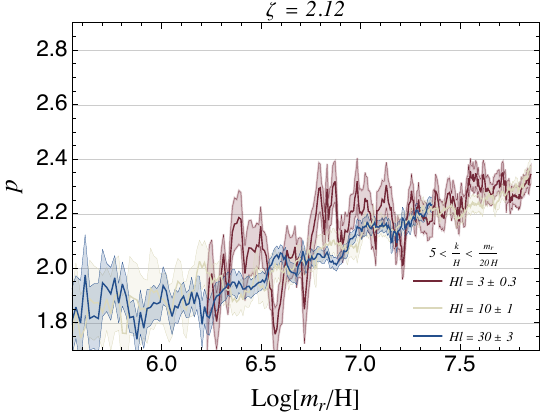}
\caption{\small The spectral index $p$ of string fluctuation for super-Hubble string of specific lengths $\ell = 3H^{-1}$, $10H^{-1}$, and $30H^{-1}$. Three panels in vertical order differ by the size of the fit interval of the rescaled momentum. The left and right panels were based on data assisted by the fat-string pre-evolution for the benchmark scenario of $\xi_0 = 0.2$ (left) and the thermal pre-evolution for the benchmark scenario of $\zeta = 2.12$ (right).
}
\label{fig:strp:4096:fixedHL}
\end{center}
\end{figure}

\begin{figure}[tp]
\begin{center}
\includegraphics[width=0.45\textwidth]{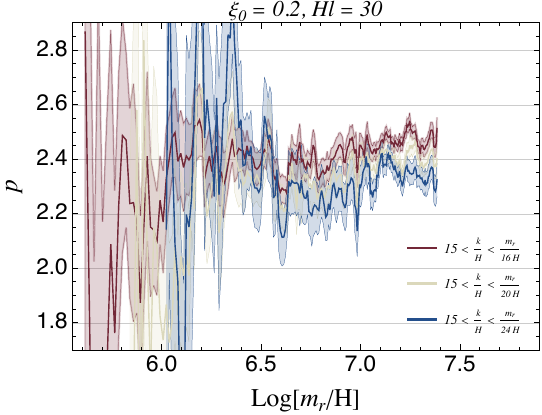} \quad
\includegraphics[width=0.45\textwidth]{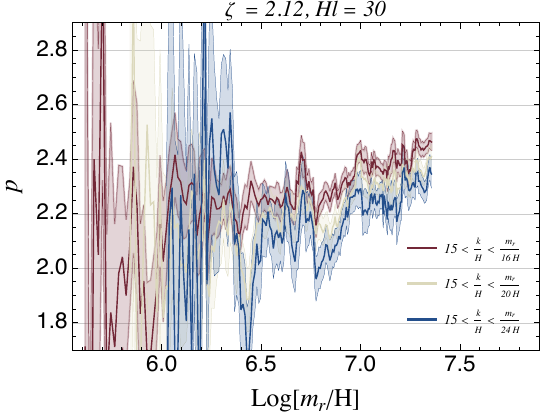}
\\[15pt]
\includegraphics[width=0.45\textwidth]{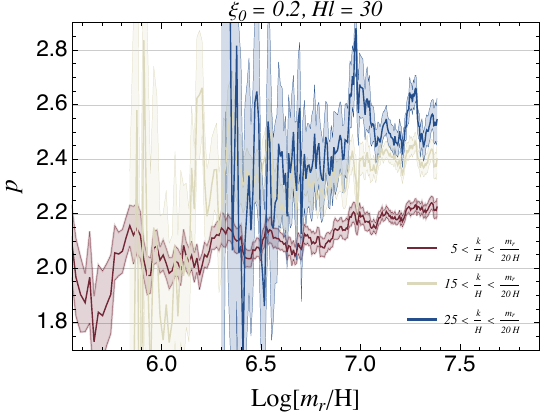} \quad
\includegraphics[width=0.45\textwidth]{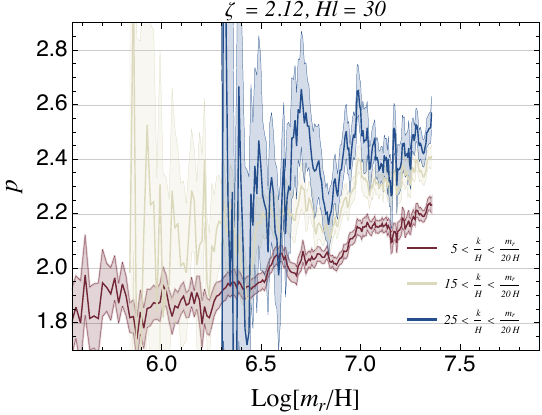}
\caption{\small The spectral index $p$ of string fluctuation for super-Hubble string of length $\ell = 30H^{-1}$, using data assisted by the fat-string pre-evolution for the benchmark scenario of $\xi_0 = 0.2$ (left) and the thermal pre-evolution for the benchmark scenario of $\zeta = 2.12$ (right).
}
\label{fig:strp:4096:fixedHL30fixed}
\end{center}
\end{figure}

We have argued that the power law fall-off behavior of string fluctuations should be similar for strings of $\ell \gg H^{-1}$, that is once strings are much longer than $H^{-1}$, since the statistical property is extracted by averaging over a large number of Hubble patches. 
This expectation is clearly confirmed in the numerical simulation presented in Fig.~\ref{fig:strp:4096:fixedHL}, where spectral indices $p$ for strings of three definite lengths exhibit similar values, for two sets of data with $\xi_0 = 0.2$ and $\zeta = 2.12$ assisted by the fat-string and thermal pre-evolutions, respectively.
Additionally, in Fig.~\ref{fig:strp:4096:fixedHL30fixed} (somewhat redundant to those in Fig.~\ref{fig:strp:4096:fixedHL}), we show the spectral index $p$ for various UV and IR cuts for super-Hubble strings of the length $\ell \sim 30H^{-1}$ for both sets of data.

\section{Cosmic strings as random walks}
\label{sec:cs:randomwalks}

At a large length scale, a string configuration is expected to follow a well-behaved random walk with a step size of the correlation length which is a Hubble length, $H^{-1}$. The string power spectrum $\mathcal{T}_n^\ell$ in this regime (a low momentum region) from lattice simulations should match to the expectation from random walks. We explicitly check this in this section. 

\begin{figure}[tp]
\begin{center}
\includegraphics[width=0.48\textwidth]{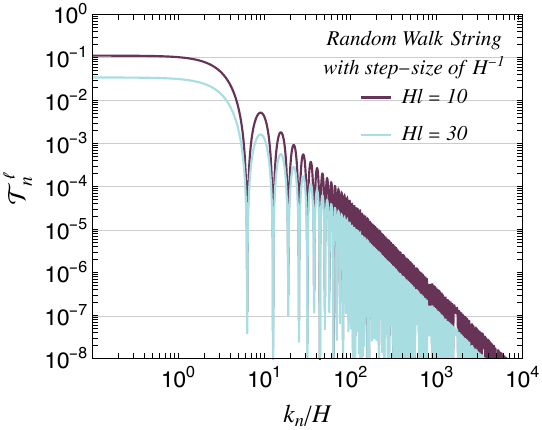}
\\[10pt]
\includegraphics[width=0.48\textwidth]{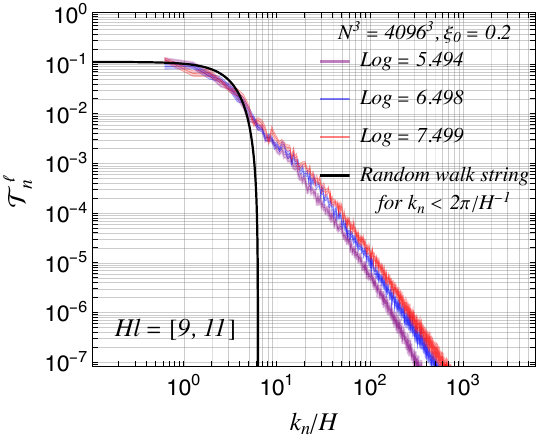}\quad
\includegraphics[width=0.48\textwidth]{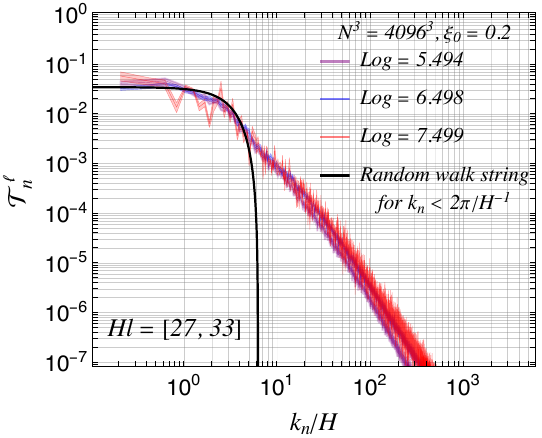}
\caption{\small The string power spectrum $\mathcal{T}_n^\ell$ of random walk strings with the step size of $H^{-1}$ and step numbers of $10$ and $30$, respectively, corresponding to string lengths of $\ell = 10 H^{-1}$ and $30 H^{-1}$.
}
\label{fig:randomWalk:4096}
\end{center}
\end{figure}

A string loop is generated from $N$ steps of random walks with a constant step size of $H^{-1}$, and it has a length of $\ell = N H^{-1}$.
After some alegbra, the power spectrum of a random walk string loop of a length $\ell$ and a step size $H^{-1}$ can be straightforwardly derived to be
\begin{equation}\label{app:randomW:string}
 \mathcal{T}_n^\ell = \frac{1}{H\ell-1} \frac{ \sin^2 \left ( \frac{k_n}{2H} \right )}{\left ( k_n/(2H) \right )^2}~,
\end{equation}
where $k_n = 2\pi n/\ell =  2\pi n/(NH^{-1})$ and it is illustrated in the top panel of Fig.~\ref{fig:randomWalk:4096}. 
The power spectrum of the random walk string in Eq.~\ref{app:randomW:string} should match to the real data only for the low momentum region $k_n < 2\pi/H^{-1}$ (or modes of $n<N$) since the random step size was set to a constant $H^{-1}$ and thus it can not account for fluctuations at a scale shorter than the Hubble length. 
The region of $k_n < 2\pi/H^{-1}$ covers the range until the first deep of Eq.~\ref{app:randomW:string} as $\frac{k_n}{2H} < \pi$.
In bottom panels of Fig.~\ref{fig:randomWalk:4096}, the power spectra of cosmic strings of lengths $\ell = 10H^{-1}$ and $30H^{-1}$ (up to 10\% widths) from lattice simulations are overlaid with the profiles of the random walk strings of step size of the Hubble length and step numbers of $10$ and $30$, respectively. As is evident in Fig.~\ref{fig:randomWalk:4096}, they match well.
Although this is not the momentum region of interest that will lead to the axion spectrum, it provides a nice validation of our analysis on data from lattice simulations.

\section{Evolution of cosmic strings}
\label{sec:dyn:string}

\subsection{Tracing causal dynamics of strings}

The causal dynamics of strings are traced in the following way. 
A string is identified as a zero locus of a complex field and is what is primarily subject to the causality is the field dynamics. Since the fluctuation of a massive radial mode around string core can be superluminal~\footnote{For instance, a simusoidal fluctuation $u(x,\, t) = \sin(kx - \omega t)$ satisfies the Klein-Gordon equation $(\partial_{tt} - \partial_{xx} + m^2 )u = 0$. The zero of $u$ at $x=\frac{\omega}{k} t$ has a velocity of $\frac{\omega}{k}$ which is superluminal when $m>0$.}, it may cause an apparent superluminal motion of a string. We define the causal connection between strings at two consecutive times up to the effect due to radial mode fluctuations within the string core region.

For given two time slices $t_1$ and $t_2$ (with $t_1 < t_2$), imagine a situation where there are $n_1$ strings at time $t_1$ and $n_2$ strings at time $t_2$. 
For each string at earlier time $t_1$, we identify a set of strings contained in the future light-cone of the corresponding string up to a margin of  $m_r^{-1}\delta_r$ for a superluminality. A fiducial value for $\delta_r$ is taken to be an order one value.
Similarly for each string at later time $t_2$, a set of strings contained in the past light-cone up to the same margin of $m_r^{-1}\delta_r$ is identified.
Five types of typical string dynamics include a single string motion, evaporation, creation, merger, and branching. A string intercommutation induces either a merger or a branching. A merger of $n$ strings into a single string or a branching of a single string into $n$ strings occurs as a result of $n-1$ intercommutations. A single string dynamics corresponds to the situation where the future or past light-cone contains only single string.

\subsection{Property of single strings}
\label{app:sec:singlestring}

\begin{figure}[tp]
\begin{center}
\includegraphics[width=0.48\textwidth]{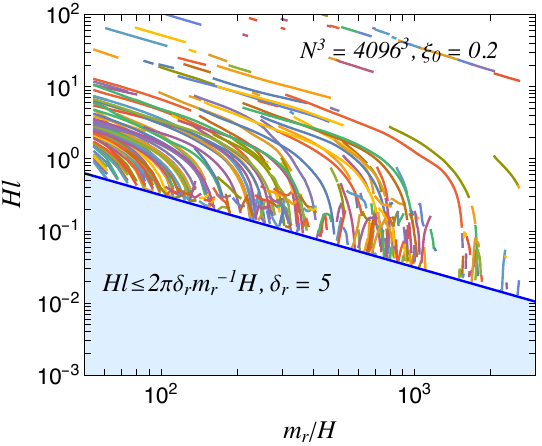}\quad
\includegraphics[width=0.48\textwidth]{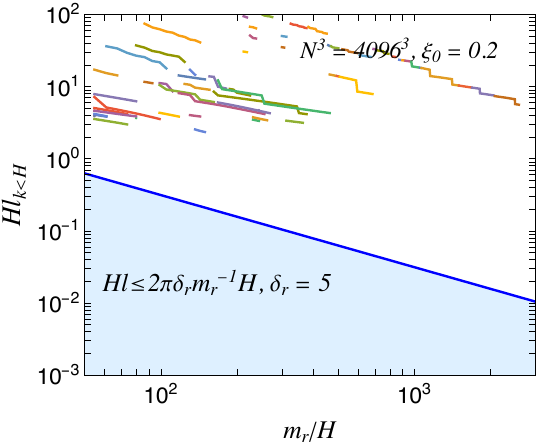}
\caption{\small The time evolutions of $H\ell$ of all single strings identified in one lattice simulation for the benchmark scenario of the initial condition $\xi_0 = 0.2$, using the fat-string pre-evolution. The contribution to the string length from modes of $k>H$ are removed in the right panel.
}
\label{fig:HLkcutoffs:4096}
\end{center}
\end{figure}

\begin{figure}[tp]
\begin{center}
\includegraphics[width=0.48\textwidth]{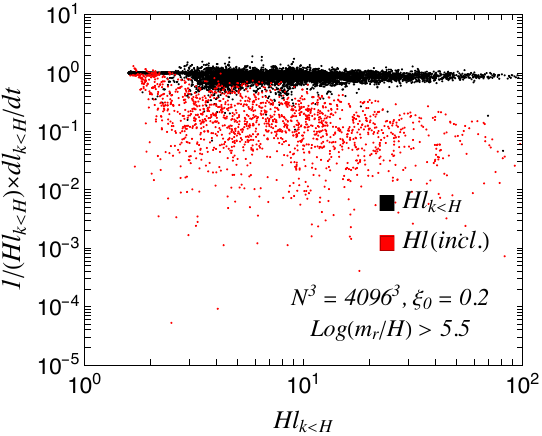}\quad
\includegraphics[width=0.48\textwidth]{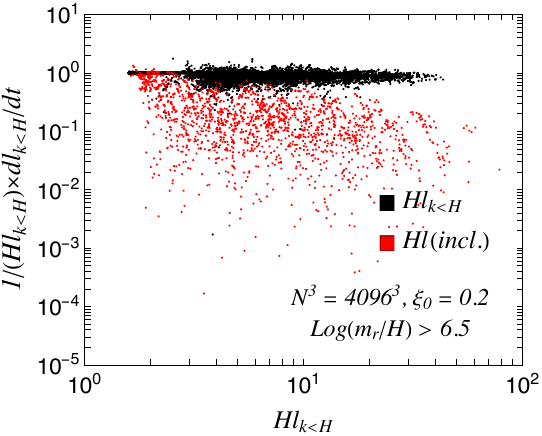}
\caption{\small The normalized differential rate of string length in time, restricted only to the low momentum $k<H$, as a function of $H\ell_{k<H}$ for two dynamic time ranges. $10^4$ randomly chosen single strings out of those from 100 independent lattice simulations were used for both plots.
}
\label{fig:HLdLdtkcutoffs:4096}
\end{center}
\end{figure}

In Fig.~\ref{fig:HLkcutoffs:4096}, time evolutions of single strings (that do not go through merging or branching), identified in an independent lattice simulation are illustrated. The solid blue lines correspond to the situation of $H\ell = 2\pi m_r^{-1} H \delta_r$ (with the choice of $\delta_r =5$) below which objects are not counted as proper strings.  As is evident in the left panel of Fig.~\ref{fig:HLkcutoffs:4096}, the lengths of strings are proportional to the blue line, or $H\ell \propto m_r^{-1}H$, which means that string lengths stay constant over time until they enter the Hubble horizon. Once they enter the Hubble horizon, the lengths drops rapidly and some bounce back from zero. We suspect that it is due to the fact that evaporations of sub-Hubble string loops tend to be accelerated. 
In the right panel of Fig.~\ref{fig:HLkcutoffs:4096}, the evolutions of lengths of single strings, including only low momentum modes of $k < H$ or long wavelength larger than $H^{-1}$, are displayed. The length of a string for $k<H$ is expected to follow the simple scaling with the scale factor $R(t)$, namely $\log(H\ell_{k<H}) \propto \log\dot{R} \propto -\frac{1}{2}\log\frac{m_r}{H}$, and this expectation looks consistent with the slope in the right panel of Fig.~\ref{fig:HLkcutoffs:4096}.
A similar scaling behavior for the string length of only long wavelength larger than $H^{-1}$ is also confirmed in Fig.~\ref{fig:HLdLdtkcutoffs:4096}.
The normalized differential rate of $\ell_{k<H}$ in time $(1/\ell_{k<H})d\ell_{k<H}/dt$ is expected to scale as $\dot{R}/R = H$. 
As is evident in Fig.~\ref{fig:HLdLdtkcutoffs:4096}, a constant scaling $1/(H\ell_{k<H})d\ell_{k<H}/dt \sim 1$ is numerically confirmed by black colored data points in Fig.~\ref{fig:HLdLdtkcutoffs:4096}. The red colored data points in Fig.~\ref{fig:HLdLdtkcutoffs:4096} are for string lengths which includes all the momenta (denoted by `incl').

In Section~\ref{sec:cs:randomwalks}, we have discussed the property of the random walk string loop whose correlation length, or the step size, is the Hubble length. The string spectra at low momenta smaller than the Hubble scale should be similar to it as was illustrated in Fig.~\ref{fig:randomWalk:4096}. The sum of line segments of a random walk string accounts for the total length of a string loop of a momentum $k \lesssim H$. The string spectra, which are relevant for the axion dark matter, are those with much smaller wavelengths than the Hubble length, and those fluctuations can be thought of being overlaid on top of line segments of a random walk string.
The ratio of the string length, restricted to $k \lesssim H$, to the length of a string is approximated as $\ell_{k\lesssim H}/\ell = 1 + \mathcal{O}(\epsilon^2)$ in the almost straight string limit. Up to an overall unspecified factor, the size of $\epsilon^2$ may be extracted from the ratio 
$\ell_{k\lesssim H}/\ell$, and it can serve the measure for the validity of our perturbative approximation.

Our simulation shows that $\ell_{k<k_\text{cutoff}}$ is proportional to $\ell$ for sufficiently long strings. The proportionality asymptotes roughly a constant at late times, or large $\log\frac{m_r}{H}$ values. We have computed the ratio for strings of $\ell H \geq 5$ for a few choices of cutoffs:
$\ell_{k\lesssim 1.5H}/\ell \sim 0.645 \pm 0.003$, $\ell_{k\lesssim 3H}/\ell \sim 0.820 \pm 0.002$ (the ratio gets reduced with a lowering cutoff and vice versa). This implies that roughly 20 $\sim$ 30\% of string length are distributed over the high momentum amplitudes. The fact that $\mathcal{O}(\epsilon^2) \sim [0.2,\, 0.3]$ implies an order of magnitude of $\epsilon$ being roughly 50\%. Since all the higher order corrections in our derivation of the axion power spectrum are an order of 
$\mathcal{O}(\epsilon^2)$ or higher, the almost straight string approximation seems valid as long as $k \gg H$ is assumed, and it becomes invalid as the momentum approaches the Hubble scale.

\newpage

{\small
\bibliography{lit}{}
\bibliographystyle{JHEP}}

\end{document}